\documentclass[a4paper, aps, prx, twocolumn]{revtex4-2}

\usepackage{braket}
\usepackage{graphicx}
\usepackage{amsmath}
\usepackage{amssymb}
\usepackage{hyperref}
\usepackage[all]{hypcap}
\hypersetup{colorlinks=True,allcolors=blue}
\usepackage{xcolor}
\usepackage{subfigure}
\usepackage{nicefrac}
\usepackage{bm}
\usepackage{bbm}
\usepackage{mathtools}  

\newcommand{\gluebra}[1]{\prescript{}{g}{\langle #1|}}

\newcommand{\vac}{\textrm{vac}}
\newcommand{\eff}{\textrm{eff}}

\newcommand{\cavcoupling}{U}

\newcommand{\WMIaff}{\affiliation{Technical University of Munich, TUM School of Natural Sciences, Physics Department, 85748 Garching, Germany}\affiliation{Walther-Meissner-Institut, Bayerische Akademie der Wissenschaften, 85748 Garching, Germany}\affiliation{ Munich Center for Quantum Science and Technology (MCQST), 80799 Munich, Germany}}

\begin{document}

\title{Cavity models of non-Abelian gauge theories}
\author{Lucia Valor}\WMIaff
\author{Jacquelin Luneau}\WMIaff
\author{Klaus Liegener}\WMIaff
\author{Stefan Filipp}\WMIaff
\author{Peter Rabl}\WMIaff

\date{\today}
\begin{abstract}
In quantum optics and related fields, cavity quantum electrodynamics provides a powerful framework for modeling the low-energy properties of bound charges coupled to electromagnetic fields, using a minimal set of atomic levels and isolated photonic modes. Here, we develop a similar framework for non-Abelian gauge theories, focusing on the pure-gauge dynamics of SU(2) fields as the conceptually simplest example. Starting from the corresponding Yang-Mills Lagrangian, we derive an SU(2) cavity Hamiltonian in the Coulomb gauge, represented in terms of only three localized bosonic modes. Unlike its non-interacting electromagnetic counterpart, this model exhibits a nontrivial excitation spectrum featuring bound composite gauge particles (`glueballs') as the lowest-energy configurations. We extend this framework to one-dimensional arrays of coupled SU(2) cavities and discuss the  driven excitation of a glueball condensate and the clustering of multi-glueball states as two elementary interaction-induced processes in these models. These examples offer a glimpse into the multitude of novel many-body phenomena that arise within this cavity SU(2) framework and we show how this physics at the interface between particle physics and quantum optics can be experimentally explored using state-of-the-art superconducting quantum circuits.
\end{abstract}
\maketitle

\section{Introduction}

Quantum electrodynamics (QED) is the prototypical example of a U(1) gauge theory and describes interactions between charged particles and the electromagnetic field at a field-theoretical level \cite{Fock1926, Dirac1927,Schwinger1948,Schwinger1949,PeskinSchroeder1995}.
In many areas of physics, however, we are only concerned with the low-energy sector of this theory for modelling the coupling of light to atoms, molecules and solids. Cavity QED \cite{haroche2013ExploringQuantumAtoms,Raimond2001,Walther2006} provides a minimal framework to study such interactions in terms of few-level emitters coupled to a single radiation mode. Despite this radical simplification, cavity QED has served as a primary workhorse for exploring elementary quantum optical phenomena as well as emerging quantum technological applications. Recent advances in nanophotonics and circuit QED~\cite{Wallraff2004,Gu2017,Blais2021}—where artificial superconducting atoms are strongly coupled to quantized microwave modes—have pushed the experimental capabilities in this field to unprecedented levels, enabling the exploration of light-matter interactions in regimes previously inaccessible in nature.  For example, coupled cavity arrays~\cite{Yariv1999,GobanNatComm2014,LodahlRMP2015,liu_quantum_2017,ChangRevModPhys2018,Ferreira2021,JouannyNatComm2025} can drastically reduce the effective speed of light~\cite{Calajo2017} and give rise to novel phases of strongly correlated photons~\cite{Hartmann2006, Greentree2006, Angelakis2007,CarusottoCiuti2013,Carusotto2020, Ma2019}. High-impedance resonators can boost the effective fine-structure constant~\cite{Devoret2007,DeBernardis2018,Kuzmin2019} to explore ultrastrong-coupling effects~\cite{Ciuti2005,FornDiaz2019,FriskKockum2019}, exotic phases of light and matter~\cite{Schuler2020,Bacciconi2025} and vacuum-induced modifications of material properties~\cite{GarciaVidal2021, Bloch2022, Schlawin2022}. Consequently, cavity QED has evolved into a versatile field, exploring a rich phenomenology of few- and many-body physics far beyond its original scope.

In this paper, we extend the framework of cavity QED to non-Abelian gauge theories~\cite{PeskinSchroeder1995,Abers1973GaugeTheories}. Our objective is to derive analogously minimal models of fundamental processes for those theories and to explore the novel few- and many-body phenomena that emerge within this description. To this end, we focus on the pure-gauge dynamics of an SU(2) gauge theory, which---unlike its Abelian counterpart---exhibits self-interacting fields even in the absence of matter. SU(2) is the gauge group of the weak interaction and, as the simplest non-Abelian group, it is also investigated as a testbed for pure-gauge processes in quantum chromodynamics, such as confinement and the formation of glueballs~\cite{PeskinSchroeder1995,Creutz1983}.
Starting from the continuum Yang-Mills Lagrangian for SU(2) \cite{PeskinSchroeder1995,YangMills1954,Abers197313YangMillsField}, we derive the corresponding cavity formulation by quantizing the gauge fields in the Coulomb gauge and restricting them to a small number of confined modes. In the single-mode limit, this approach yields a minimal cavity SU(2) Hamiltonian, wherein three bosonic modes of different gauge flavors are coupled via density-density and two-particle exchange interactions. Despite its self-interacting nature, the underlying SU(2) symmetry permits an exact characterization of the spectrum of this cavity Hamiltonian. From the corresponding eigenstates, we identify composite glueball states~\cite{Glueballs} 
as the energetically lowest multi-boson configurations. As an illustrative example, we discuss the excitation of a glueball condensate by an external driving field and show that the resulting non-Abelian dynamics of even a single SU(2) cavity is already surprisingly rich and differs in many aspects from the behavior expected from driven photonic modes in conventional cavity QED.

By retaining additional modes in the expansion of the gauge fields, the cavity SU(2) Hamiltonian can be systematically refined to approach the original continuum limit by resolving smaller and smaller spatial structures within the cavity volume.
In this sense,  our framework is complementary to lattice gauge theories 
\cite{Wilson1974,KogutSusskind1975,Creutz1983,Kogut1983}, where global properties are typically captured `bottom-up' by increasing the number of elementary vertices or plaquettes. Compared to other studies of gauge theories in spatially confined regions~\cite{tHooft1979,Bjorken1979, Luscher1983, LuscherMunster1984,
vanBaalKoller1987,vanBaal2000}---which have recently regained interest in the context of quantum simulation~\cite{ButtDraperShen2023,HartnettLiaoRinaldi2026,Bergner2026}---the present approach introduces an additional level of abstraction by expanding the field in a small number of finite-frequency modes. Most importantly, this simplification establishes a direct correspondence to the field of modern cavity QED and enables us to transfer many engineering and control concepts from this field to non-Abelian gauge theories. As a basic example, we extend our theory to a one-dimensional (1D) array of SU(2) cavities, where the gauge bosons can tunnel between neighboring cavity sites and propagate through the array.  This setup provides a minimal configuration to explore the competition between the kinetic and interaction energies of non-Abelian gauge bosons, and we discuss the formation of bound multi-glueball states in the interaction-dominated regime.

Since non-Abelian gauge bosons associated with the fundamental forces can only be observed indirectly in experiments, a crucial practical advantage of our cavity formulation of SU(2) is that it renders the resulting models accessible to near-term quantum simulation platforms. Models similar to those derived here already appear naturally in the context of spinor condensates of ultracold atomic gases~\cite{Law1998,Ho1998,Koashi2000,KawaguchiUeda2012,DemlerZhou2002,Imambekov2003,Rizzi2005}, and could be further adapted to achieve exact formal matching. Here, we propose a circuit QED implementation of SU(2) cavities, which in the long-run can leverage the exquisite engineering and quantum-control capabilities of this platform~\cite{Blais2021,Krantz2019}. In this implementation, the gauge bosons are represented by three quantized microwave resonators, while the necessary interaction terms can be engineered with the help of nonlinear Josephson elements. Compared to the vast body of quantum simulation schemes for lattice gauge theories~\cite{Banerjee2013,ZoharCiracReznik2013, Tagliacozzo2013,Marcos2013,Marcos2014,Mezzacapo2015,Brennen2016,ZoharCiracReznik2016,Banuls2020,KlcoStrykerSavage2020,Wiese2021,Rahman2021,Atas2021,GonzalezCuadra2022,Zhang2023,Bauer2023,HalimehSpinExchange2024,Calajo2024,Crane2024,Halimeh2025,Busnaina2025,John2026,Ilcic2026,Liegener2026VQE,AlcaineCuervo2026},
this approach naturally maps the gauge fields onto continuous-variable degrees of freedom for extended modes, rather than relying on finite-dimensional representations of link variables. Consequently, a significantly larger Hilbert space is accessible for a given number of circuit elements. This feature makes cavity SU(2) models particularly interesting for near-term quantum simulation applications, but also extends the range of effective models and physical phenomena that can be explored in this context. 

The remainder of the paper is structured as follows. First, in Sec.~\ref{sec:su2_cavity}, we derive a cavity formulation of a pure SU(2) Yang--Mills theory by quantizing the gauge fields in the Coulomb gauge and restricting them to a single confined mode. In Sec.~\ref{sec:Spectrum_of_SU2} we discuss the spectrum of this cavity Hamiltonian and identify the composite glueball states as a family of low-energy excitations. In Sec.~\ref{sec:DrivenCavity}, we show how external drives can be used to excite the otherwise trivial vacuum into a condensate of such glueball states or other composite eigenstates. In Sec.~\ref{sec:waveguide_su2}, we extend the model to a 1D array of coupled cavities, where perturbative hopping processes generate effective glueball motion and attractive interactions, leading to bound two- and few-glueball states. Finally, in Sec.~\ref{sec:analog_simulation}, we discuss a potential implementation of the described model with superconducting circuits and summarize our main findings in Sec.~\ref{sec:conclusion}.

\begin{figure}
    \centering
    \includegraphics[width=\linewidth]{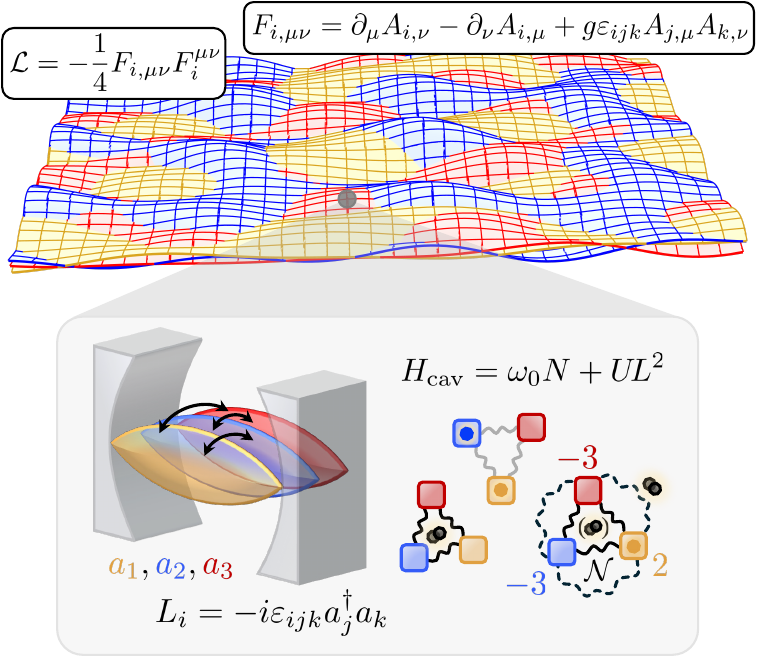}
    \caption{Illustration of a non-Abelian gauge theory with continuous gauge fields $A_{i=1,2,3}$ described by a Yang-Mills Lagrangian. With the framework of cavity SU(2), the gauge fields inside a confined region in space are represented by a single localized mode function. The resulting reduced cavity Hamiltonian $H_{\rm cav}$ describes three interacting bosonic modes with annihilation (creation) operators $a_i$ ($a_i^\dag$) and supports different types of bound and excited glueball states that are analyzed in this work. See text for more details. }
    \label{fig:overview}
\end{figure}

\section{SU(2) cavity gauge theory}\label{sec:su2_cavity}
We are interested in the pure-gauge dynamics of a non-Abelian gauge theory, as described by the Yang-Mills Lagrangian~\cite{PeskinSchroeder1995,YangMills1954,Abers197313YangMillsField}
\begin{equation}\label{eq:YMLagragian}
    \mathcal{L} = -\dfrac{1}{4}F_{i,\mu\nu}F_i^{\mu\nu}.
\end{equation}
Here $\mu,\nu$ are the spacetime indices and $i$ labels the different flavors of gauge potentials. We adopt the convention that repeated indices are implicitly summed over. In terms of the components $A_i^\mu=(A_i^0, A_i^1,A_i^2,A_i^3)\equiv(A_i^0,\vec A_i)$ of the gauge potentials, the field tensor is given by
\begin{equation}\label{eq:F}
 F_{i}^{\mu\nu} =\partial^\mu A_{i}^{\nu}-\partial^\nu A_{i}^{\mu}+g\varepsilon_{ijk}A_{j}^{\mu} A_{k}^{\nu},  
\end{equation}
where $g$ is the coupling constant and $\varepsilon_{ijk}$ are the structure constants of the gauge group.  For the case of SU(2), we obtain three gauge fields with $i={1,2,3}$ and $\varepsilon_{ijk}$ is the Levi-Civita symbol. The nonlinear term proportional to~$g$ in Eq.~\eqref{eq:F} represents the non-Abelian nature of the theory, in which the gauge fields interact with themselves, even in the absence of matter.  

Note that for the sake of simplicity, we work in dimensionless units. We also follow the usual convention with $\mu=0$ labeling the time component and Latin indices labeling the spatial components of the four-vectors. For raising and lowering of indices, we assume that the metric is  Lorentzian with signature $(+,-,-,-)$.

\subsection{Canonical Hamiltonian and the Gauss law}
Starting from the Lagrangian in Eq.~\eqref{eq:YMLagragian}, we quantize the theory in the canonical Hamiltonian framework by treating the spatial components of the gauge fields $A_{i}^a$ and their conjugate momenta, 
\begin{equation}
    \Pi^a_{i} =\dfrac{\partial\mathcal{L}}{\partial(\partial^0 A_{i}^a)} = -F_{i,0a} \equiv -E_i^a,
\end{equation}
as canonical variables. After a Legendre transformation, we obtain the Hamiltonian density $\mathcal{H}=\Pi^a_i  \dot A_{i}^a-\mathcal{L}$, which can be written as
\begin{equation}
 \begin{aligned}\label{eq:H_start}
   \mathcal{H}&=\dfrac{1}{2} E_{i}^a E_{i}^a + \dfrac{1}{4} F_{i,ab} F_{i}^{ab} + A^0_{i}G_i. 
\end{aligned}   
\end{equation}
The first two terms in this equation represent the non-Abelian generalizations of the electric and magnetic energy densities, respectively, where the magnetic term $\propto F_{i,ab}F_{i}^{ab}$ contains both the curl of the gauge fields and the self-interaction. The last term in Eq.~\eqref{eq:H_start}, where  
\begin{equation}\label{eq:G}
G_i=-\partial_aE_{i}^a+g\varepsilon_{ijk}A_j^{a}E_k^{a},
\end{equation}
involves the temporal components $A^0_i$. Since their associated canonical momenta vanish ($E^0_i=0$), these components are non-dynamical degrees of freedom of the theory. To ensure that this property holds for all times, we must impose the additional dynamical constraints 
\begin{equation}\label{eq:Gauss}
\dot{E}^0_{i}=\{E^0_{i},H\}=G_i=0.
\end{equation}
These constraints represent the non-Abelian generalization of the usual Gauss law in electrodynamics and must be satisfied by physically meaningful configurations of gauge fields at all times.

\subsection{Transverse and longitudinal fields}

In QED, in the absence of charges, the Gauss law in Eq.~\eqref{eq:Gauss} reduces to $\partial_aE^a=0$, which can be readily satisfied by working in the Coulomb gauge with $\partial_a A^{a}=0$.  The Coulomb gauge is thus commonly used in non-relativistic scenarios to perform the quantization of the remaining transverse components of the free electromagnetic field,  $\vec A^\perp$ and $\vec E^\perp$. In this gauge, the longitudinal electrical field, $\vec E^\parallel$, is a non-dynamical degree of freedom and fixed by the Coulomb law when charges are present. 

When working in the Coulomb gauge of a non-Abelian gauge theory, $\partial_a A_{i}^a=0$, the Gauss law in Eq.~\eqref{eq:Gauss} is not immediately satisfied. Effectively, the term $\propto g$ in Eq.~\eqref{eq:G} can be interpreted as a nonvanishing charge density $\rho_i\neq 0$, which is associated with the fields themselves. More precisely, in the Coulomb gauge with $\vec A_i=\vec A^\perp_{i}$ and decomposing the electric fields into transverse and longitudinal components ($\vec E_i = \vec E^\perp_{i} + \vec E^\parallel_{i}$), the Gauss law reads
\begin{equation}\label{eq:Elong_Gausslaw}
   - \vec\nabla  \cdot\vec E^\parallel_{i}+g\varepsilon_{ijk}\vec A_{j}\cdot (\vec E^\parallel_{k}+\vec E^\perp_{k})=0.
\end{equation}
This set of equations determines the components of the longitudinal fields $\vec E^\parallel_{i}$ in terms of the transverse fields $\vec A_{j}$ and $\vec E^\perp_{k}$ and allows us, in principle, to express all fields in terms of dynamical degrees of freedom only. However, in contrast to QED, there is generally no closed solution to this set of equations. 

To overcome this problem, here we focus on weakly coupled theories, in which case Eq.~\eqref{eq:Elong_Gausslaw} can be solved iteratively in powers of $g$. This procedure is detailed in Appendix \ref{sec:Elong_derivation}, following the general approach outlined in Ref.~\cite{Abers197313YangMillsField}. By restricting this analysis to the lowest nonvanishing order, we obtain 
\begin{equation}
\vec E^\parallel_{i}(x)= - g \vec\nabla \int d^3y \; \dfrac{\rho_i(\vec y)}{4\pi|\vec x-\vec y|}+\mathcal{O}(g^2),\label{eq:f_order_g}
\end{equation}
where we have introduced the effective charge density
\begin{equation}
\rho_i = \varepsilon_{ijk}\vec A_{j} \cdot \vec E^\perp_{k}.
\end{equation}
Therefore, to lowest order in the coupling constant, we obtain the intuitive result, where the longitudinal fields follow a Coulomb field configuration with a charge density that is quadratic in the dynamical fields. 
The Hamiltonian in Eq.~\eqref{eq:H_start} can then be re-expressed in the form
\begin{equation}\label{eq:H_EBE}
    H = \int d^3x \, \mathcal{H} = \frac{1}{2} \int d^3x \left(\vec{E}_i^{\perp}\cdot\vec{E}_i^{\perp}+\vec{E}^{\parallel}_i\cdot\vec{E}^{\parallel}_i +\vec{B}^2\right),
\end{equation}
where $\frac{1}{2}\vec{B}^2\equiv \frac{1}{4}F_{i,ab}F_{i}^{ab}$ is the magnetic energy density. In this form, we can now insert our solution for $\vec E^\parallel_{i}\equiv \vec E^\parallel_{i}(\vec A_i,\vec E^\perp_{i})$ in Eq.~\eqref{eq:f_order_g} into Eq.~\eqref{eq:H_EBE} to arrive at a Hamiltonian that is expressed in terms of dynamical degrees of freedom only. 

\subsection{SU(2) Hamiltonian in the Coulomb gauge}

Once the longitudinal electric field is eliminated, the Hamiltonian density can be written entirely in terms of the transverse fields, $\mathcal{H}\equiv \mathcal{H}(\vec A_i, \vec E^\perp_{i})$.  This yields the natural decomposition $\mathcal{H}=\mathcal{H}_0+\mathcal{H}_{\mathrm{int}}$, where $\mathcal{H}_0$ contains the free, quadratic terms and $\mathcal{H}_{\mathrm{int}}$ contains all interactions. The free part,
\begin{equation}\label{eq:Hfree}
    \mathcal{H}_0 =\frac{1}{2}\!\left[\vec{E}_i^{\perp}\cdot\vec{E}_i^{\perp}
    +\left( \vec \nabla \times \vec{A}_i\right)\cdot\left( \vec \nabla \times \vec{A}_i\right)\right],
\end{equation}
is the $g=0$ limit of the Yang-Mills Hamiltonian written in the Coulomb gauge. It plays the same role as the quadratic Maxwell Hamiltonian in QED and therefore provides the natural starting point for defining the mode expansion of the gauge fields. The interaction part can be expanded perturbatively in~$g$, following the analysis in the previous section. It reads
\begin{equation}\label{eq:startH}
\begin{aligned}
\mathcal{H}_{\mathrm{int}}
= \,& -g \varepsilon_{ijk} (\partial_a A_{i}^b) A_{j}^a A_{k}^b
+\frac{g^2}{4}\left(\varepsilon_{ijk}A_{j}^a A_{k}^b\right)^2\\
&+\frac{g^2}{2}\int d^3y\,\frac{\rho_i(\vec x)\rho_i(\vec y)}{4\pi|\vec x-\vec y|}
+\mathcal{O}(g^3).
\end{aligned}
\end{equation}
The interacting Hamiltonian collects all non-Abelian contributions: the cubic and quartic self-interaction terms arising from~$F_{i,ab}F_i^{ab}$ and the non-local Coulomb-like interaction generated by solving Gauss' law. These terms couple the free transverse modes and encode the non-Abelian dynamics.

\subsection{A cavity theory for SU(2)}

The Yang-Mills Hamiltonian derived above in Eq.~\eqref{eq:Hfree} and Eq.~\eqref{eq:startH} is formulated in terms of continuous fields. In a next step, our goal is to derive a cavity model of SU(2), where, in close analogy to cavity QED, the gauge fields are confined in a small region of space and expressed in terms of a set of discretely spaced modes. To do so, we note that the free Hamiltonian given in Eq.~\eqref{eq:Hfree} has the same quadratic structure as the Maxwell Hamiltonian, and can thus be diagonalized and quantized in terms of the same set of eigenmodes with normalized, transverse mode functions $\vec u_{n}(\vec x)$ and frequencies $\omega_n$. These modes satisfy the Helmholtz equation 
\begin{equation}
(\vec \nabla^2 + \omega_n^2)\vec u_n=0, 
\end{equation}
where in cavity QED the boundary conditions would be determined, for example, by the cavity mirrors. Here we do not make reference to any specific physical settings, but simply assume that the gauge fields are constrained to a small volume $V$, for which such a mode expansion exists. In terms of these modes, the quantized gauge fields can then be expressed as 
\begin{gather}
\vec A_i(\vec x)=\sum_n \sqrt{\dfrac{1}{2\omega_n}}\; \big[\vec u_{n}(\vec x) a_{i,n}+\vec u^{*}_{n}(\vec x) a^\dag_{i,n}\big],\label{eq:mode_exp_A}\\
\vec E_{i}^\perp(\vec x)= i \sum_n\sqrt{\dfrac{\omega_n}{2}}\; \big[\vec u_{n}(\vec x) a_{i,n}-\vec u^{*}_{n}(\vec x) a_{i,n}^{\dagger}\big],
\label{eq:mode_exp_E}
\end{gather}
where the operators $a_{i,n}$ ($a^\dag_{i,n}$) annihilate (create) gauge bosons in mode $n$ with SU(2) flavor index $i$. They satisfy the bosonic commutation relations  $[a_{i,n},a^\dag_{j,m}]=\delta_{ij}\delta_{nm}$. In terms of these eigenmodes, the free Hamiltonian is diagonal,
\begin{align}\label{eq:H0_singlemode}
    H_0 =\int_V d^3x~\mathcal{H}_0(x) = \sum_{i,n}\omega_n a_{i,n}^{\dagger}a_{i,n},
\end{align}
and simply describes three sets of independent bosonic modes. Note that in Eq.~\eqref{eq:H0_singlemode} we have already dropped the vacuum energy.

The key difference to QED arises from the interaction term $\mathcal{H}_{\rm int}$ given in Eq.~\eqref{eq:startH}, which we can also expand in terms of uncoupled eigenmodes. We obtain a general normal-ordered expression of the form 
\begin{align}
H_{\rm int}=\sum_{\substack{i\\n,m}}
\left[
G_{\substack{i\\nm}}^{(1,1)}a_{i,n}^{\dagger}a_{i,m}
+\left(G_{\substack{i\\nm}}^{(0,2)}a_{i,n}a_{i,m}+\text{H.c.}\right)
\right]&\nonumber\\
+\sum_{\substack{i,j,k\\n,m,p}}
\left[
\left(G_{\substack{ijk\\nmp}}^{(0,3)}a_{i,n}+G_{\substack{ijk\\nmp}}^{(1,2)}a_{i,n}^{\dagger}\right)a_{j,m}a_{k,p}
+\text{H.c.}\right]&\nonumber\\[4pt]
+\sum_{\substack{i,j,k,l\\n,m,p,q}}
\left[\left(G_{\substack{ijkl\\nmpq}}^{(0,4)}a_{i,n} + G_{\substack{ijkl\\nmpq}}^{(1,3)}a_{i,n}^{\dagger}\right)
a_{j,m}a_{k,p}a_{l,q}
+\text{H.c.}\right]&\nonumber\\[4pt]
+\sum_{\substack{i,j,k,l\\n,m,p,q}}
G_{\substack{ijkl\\nmpq}}^{(2,2)}
a_{i,n}^{\dagger}a_{j,m}^{\dagger}a_{k,p}a_{l,q}.\hspace{2.85cm} &
\label{eq:Hint_general_exp}
\end{align}
It describes various three-photon and four-photon processes, which arise from the expansion of the different terms in $\mathcal{H}_{\rm int}$. The corresponding coupling constants depend on overlap integrals over different mode functions~$\vec u_n(\vec x)$ and they are summarized in Appendix~\ref{sec:mode_expansion}. 

Note that, at the classical level, the product terms in Hamiltonian $\mathcal{H}_{\rm int}$ in Eq.~\eqref{eq:startH} can be commuted indifferently, but this can lead to different normal-ordered Hamiltonians at the quantum level. The first line of Eq.~\eqref{eq:startH} involves only commuting fields~$A_{i}^a$, so they can be unambiguously substituted by the corresponding operators in Eq.~\eqref{eq:mode_exp_A}. In the expression for the Coulomb interaction in the second line of Eq.~\eqref{eq:startH}, the fields contributing to the effective charge, $\rho_i=\varepsilon_{ijk}A_{j}^a E_{k}^a$, have different flavors, so they commute and can be substituted unambiguously. However, the product of charges $\rho_i\rho_i$ represents a product of four fields that are not mutually commuting. In Eq.~\eqref{eq:Hint_general_exp}, we have used the natural choice where the fields are replaced by their quantized operators in each source $\rho_i$ separately and multiplied afterwards. 
This choice results in a Hermitian $H_{\rm int}$, and any other ordering of the four fields either breaks Hermiticity or reproduces Eq.~\eqref{eq:Hint_general_exp} up to a constant. The normal ordering of the quartic terms induces the additional quadratic terms in the first line of Eq.~\eqref{eq:Hint_general_exp}, while that of the cubic term induces no linear corrections. We therefore conclude that the ambiguity of operator ordering~\cite{christ1980OperatorOrderingFeynman, greensite2020IntroductionConfinementProblem} has no influence on the overall structure of $H_{\rm int}$.

\subsection{Single-mode approximation}

In general, the interaction Hamiltonian $H_{\rm int}$ couples all modes. However, by assuming that the coupling constants~$G\sim g,g^2$ are small compared to the typical energy spacings $|\omega_n-\omega_m|$ and that there are no other accidental degeneracies, the dynamics of the gauge fields can be restricted to a single mode, which we label by the index $n=0$ in the following.  This mode can represent the fundamental mode, but also any other higher energy mode in which the system is initialized.
Under this single-mode approximation and by identifying $a_i \equiv a_{i,0}$, we obtain
\begin{equation}\label{eq:field_restriction_single_mode}
\begin{split}
    \vec A_i(x) &= 
    \sqrt{\frac{1}{2\omega_0}}\, \left[ \vec u_{0}(x)\, a_i + \vec u^*_{0}(x)\, a_i^\dag\right],\\
        \vec E_i^\perp(x) &= 
    i \sqrt{\frac{\omega_0}{2}}\, \left[ \vec u_{0}(x)\, a_i -  \vec u^*_{0}(x)\, a_i^{\dag}\right].
\end{split}
\end{equation}
Under the same approximation, the total Hamiltonian $H=H_0+H_{\rm int}$ reduces to the cavity SU(2) Hamiltonian 
\begin{align}\label{eq:H_singleMode_expanded}
  H_{\rm cav} = \sum_i (\omega_0 +2U) n_i  +\cavcoupling\sum_{i\neq j} \left(n_in_j - a_i^{\dagger\,2} a_j^2 \right),
\end{align}
where $n_i=a_i^\dagger a_i$. The interaction energy, 
\begin{align}
\cavcoupling = \cavcoupling_4 + \cavcoupling_C > 0,
\end{align}
contains two contributions, 
\begin{align}
    \cavcoupling_4 &= \frac{g^2}{8\omega_0^2}\int d^3x\left(|\vec u_0(\vec x)|^4 - |\vec u_0(\vec x)\cdot\vec u_0(\vec x)|^2\right), \label{eq:U4}\\
    \cavcoupling_C &= \frac{g^2}{8\pi}\int d^3xd^3y\, \frac{|\vec u_0(\vec x)|^2|\vec u_0(\vec y)|^2}{|\vec x-\vec y|},\label{eq:UC}
\end{align}
which arise from the direct and the Coulomb interaction between the gauge potentials, respectively. 

We see that the self-interaction between the gauge fields, either directly or via the Coulomb interaction, results in a repulsive cross-Kerr coupling between different flavors  and a two-boson hopping term. The contribution  $U_4$ is only nonvanishing for mode functions with a complex polarization vector, i.e., for modes with nonzero helicity. For the assumed expansion into a single mode, all contributions involving three gauge fields vanish exactly. For more details on the derivation of $H_{\rm cav}$, see  Appendix~\ref{sec:mode_expansion}.

As a concrete example, we assume a cubic volume $V=\ell_c^3$ with periodic boundary conditions and consider the mode function 
\begin{equation}
\vec u_0(\vec x)= \vec e_+\, \frac{1}{\sqrt{V}} \,  e^{i \vec k \cdot \vec x} ,
\end{equation}
where $\vec e_+=(1,i,0)/\sqrt{2}$ and $\vec k=(0,0,\omega_0)$. In this case we obtain
\begin{equation}
 U_4 = \frac{g^2}{8\omega_0^2 V}, \qquad \cavcoupling_C \simeq 0.075 \times \frac{g^2}{\ell_c},
\end{equation}
where we have evaluated the Coulomb integral following Ref.~\cite{ciftja2011ChargedCube}. 
This estimate shows how the two contributions to the interaction scale with the system parameters. For the validity of the single-mode approximation, we require that both interaction scales are small compared to the typical mode spacing, $\Delta \omega$. For example, for the fundamental mode with $\omega_0=(2\pi/\ell_c)=\Delta \omega$, this condition is satisfied for both terms when $g^2/(\omega_0 \ell_c)\ll 1 $. As shown in Appendix~\ref{sec:Elong_derivation}, this condition then also justifies the truncation of the Coulomb interaction to the lowest nontrivial term, as assumed in Eq.~\eqref{eq:f_order_g}, and the different approximations involved in the derivation of $H_{\rm cav}$ are mutually consistent.

\section{Spectrum of cavity SU(2)}\label{sec:Spectrum_of_SU2}
In contrast to cavity QED, where, in the absence of matter, one is left with a single harmonic oscillator, the same minimal model in SU(2) leaves us with three interacting modes, corresponding to the three SU(2) flavors. The SU(2) structure that underlies this Hamiltonian becomes apparent when introducing the operators~\cite{Law1998}
\begin{equation}\label{eq:angular_momentum_def}
    L_i = -i\,\varepsilon_{ijk}\, a_j^{\dagger } a_k.
\end{equation}
These operators couple bosons of different flavor and satisfy the SU(2) (or angular momentum) commutation relations
$[L_i,L_j] = i\,\varepsilon_{ijk}L_k$.
Using this notation, the single-mode Hamiltonian in Eq.~\eqref{eq:H_singleMode_expanded} can be re-written in a compact form as
\begin{equation}\label{eq:Ham_cavity_J}
    H_{\rm cav} = \omega_0 N + \cavcoupling  L^2,
\end{equation}
where $ L^2 = \sum_{i} L_i^2$ is the squared angular momentum operator and $N = \sum_{i=1}^{3} a_i^\dagger a_i$ the total number of gauge bosons. Note that in the following discussion we will use---depending on the context---the symbol $N$ for both the number operator and for its eigenvalues.

\subsection{Energy ladder and SU(2) glueballs} 

By expressing $H_{\rm cav}$ in terms of the total angular momentum operator and since $[N,L^2] =0$, it immediately follows that the single-mode cavity SU(2) model can be diagonalized exactly and that eigenstates can be labeled by three quantum numbers as $\ket{N,l,m}$. Here $N$ specifies the total number of excitations, $l$ is the angular-momentum quantum number, and  $m$ is the projection quantum number with $-l\leq m\leq l$. The corresponding eigenvalues are
\begin{equation}\label{eq:SpectrumExact} 
    E_{N,l} = \omega_0N +\cavcoupling l(l+1),
\end{equation} 
and they are $(2l+1)$-fold degenerate. In analogy to the choice of a quantization axis for a system with spin or orbital angular momentum,  the projection quantum number~$m$ is the eigenvalue of a chosen component of the angular momentum, for example, $L_3$, such that 
\begin{align}
    L_3\ket{N,l,m}=m\ket{N,l,m}.
\end{align}
As shown below, these three quantum numbers fully specify the eigenstate and no other multiplicity index is required. The quantum number $l$ takes all integer values smaller than or equal to $N$ and of the same parity as~$N$.

To derive these properties as well as to obtain the expression of the eigenstates in the Fock basis, it is useful to introduce the operator
\begin{align}
    B^\dagger = \frac{1}{\sqrt{6}}\left(a_1^{\dagger2} + a_2^{\dagger2} + a_3^{\dagger2}\right),
\end{align}
which creates two bosons equally distributed between the three gauge flavors. 
From an explicit calculation, one finds that $B^\dagger$ commutes with all angular momentum components
\begin{align}\label{eq:Bdag_comm_J}
    [B^\dagger,L_i] = [B^\dagger, L^2]=0.
\end{align}
This implies that, starting in the vacuum state $|{\rm vac}\rangle= |N=0,l=0,m=0\rangle$, the repeated action of $B^\dagger$ creates states of vanishing angular momentum, i.e., SU(2) singlets. In the following we  denote those states by
\begin{align}\label{eq:glueball_def}
    \ket{\mathcal{N}}_g \equiv \ket{N=2\mathcal{N},l=0,m=0}=\frac{(B^\dagger)^{\mathcal{N}}}{\sqrt{C_\mathcal{N}}} \ket{\vac}.
\end{align}
Note, however, that $B$ and $B^\dagger$ are not bosonic annihilation and creation operators.  Instead, they satisfy the commutation relation 
\begin{equation}
[B,B^\dagger]=1+\frac{2}{3} N,
\end{equation}
which affects the normalization factors $1/\sqrt{C_\mathcal{N}}$ of the states. Their values are given in Appendix~\ref{sec:supp:singlemode_algebraic} together with other useful algebraic relations between the operators~$B^\dagger$, $L_i$ and $a_i^\dagger$.

The singlet sector with $l=0$ is special because it corresponds to the non-degenerate ground state within a fixed excitation-number sector. According to Eq.~\eqref{eq:SpectrumExact}, those states have a vanishing interaction energy and
\begin{equation}
H_{\rm cav} |\mathcal{N}\rangle_g = 2\omega_0 \mathcal{N} \ket{\mathcal{N}}_g.
\end{equation} 
Since $\cavcoupling>0$, those states have a lower energy than any other configuration of $2\mathcal{N}$ gauge bosons, i.e., they are bound by the interaction. Therefore, we refer to the states $\ket{\mathcal{N}}_g$  as $\mathcal{N}$-glueball states and to $B^\dagger$ as the glueball ladder operator~\cite{Glueballs}.

\subsection{Arbitrary energy eigenstates} 

The glueball ladder operator~$B^\dagger$, together with standard properties of angular momentum operators, enables us to express an arbitrary eigenstate~$\ket{N,l,m}$ in the Fock basis of the three original bosonic modes. To do so, it is useful to introduce the modes
\begin{align}
    a_{\pm}^\dagger = \frac{1}{\sqrt{2}}\left(a_1^\dagger \pm ia_2^\dagger\right),
\end{align}
whose number operators diagonalize the angular momentum component along the quantization axis,
\begin{align}
    L_3 = a_+^\dagger a_+ - a_-^\dagger a_-.
\end{align}
As a consequence, in the sector of fixed excitation number~$N$, the angular momentum projection satisfies $|m|\leq N$, and so does the angular momentum quantum number~$l\leq N$. In particular, the maximal value $l=m=N$ is obtained with all bosons in mode $a_+$, and thus 
\begin{align}
    \ket{N,l=N,m=N} = \frac{1}{\sqrt{N!}}\left(a_+^\dagger\right)^N \ket{\vac}.
\end{align}
For a fixed $l=N$, the states with lower projection numbers, $ m< N$, are obtained by applying the angular momentum lowering operator
\begin{align}
    L_-=L_1-iL_2=\sqrt{2}\left(a_-^\dagger a_3 - a_3^\dagger a_+\right),
\end{align}
which decreases~$m$ by one. Finally, due to the commutation relation in Eq.~\eqref{eq:Bdag_comm_J}, by acting with the glueball ladder operator~$B^\dagger$ on any state does not change~$l$ and~$m$, but increases~$N$ by two. Therefore, for any integers~$\mathcal{N},l,m$, we obtain an expression for the eigenstate
\begin{align}\label{eq:general_eigenstate_exp}
    \ket{N=l+2\mathcal{N},l,m}\propto (B^\dagger)^\mathcal{N} L_-^{l-m}(a_+^\dagger)^l\ket{\vac}.
\end{align}
The corresponding normalization prefactor and ladder expressions of $L_-$ and $B^\dagger$ are given in Appendix~\ref{sec:supp:singlemode_algebraic}. Equivalently, the same eigenstate can be expressed as $\propto(B^\dagger)^\mathcal{N} L_+^{l+m}(a_-^\dagger)^l\ket{\vac}$, i.e., by starting from the $m=-l$ eigenstate and applying the raising operator $L_+=L_1+iL_2$.

This construction provides all the quantum numbers $l\leq N$ with $l$ of the same parity as $N$. Their $2l+1$ multiplicity leads to a total of $(N+1)(N+2)/2$ different states, which matches the dimension of the $N$-excitation subspace of the three-mode system. As such, they form a complete eigenbasis and no other values of $l$ are allowed. Note that a similar analysis of the angular momentum eigenstates of three bosonic modes has been previously carried out in the context of spinor condensates~\cite{Law1998,Ho1998,Koashi2000,KawaguchiUeda2012}, where interactions of the form $UL^2$ arise naturally from the symmetry of bosonic interactions.

\subsection{Flavor symmetric low-energy  states}\label{sec:flavor_symm_states}

\begin{figure*}
    \centering
    \includegraphics[width=\linewidth]{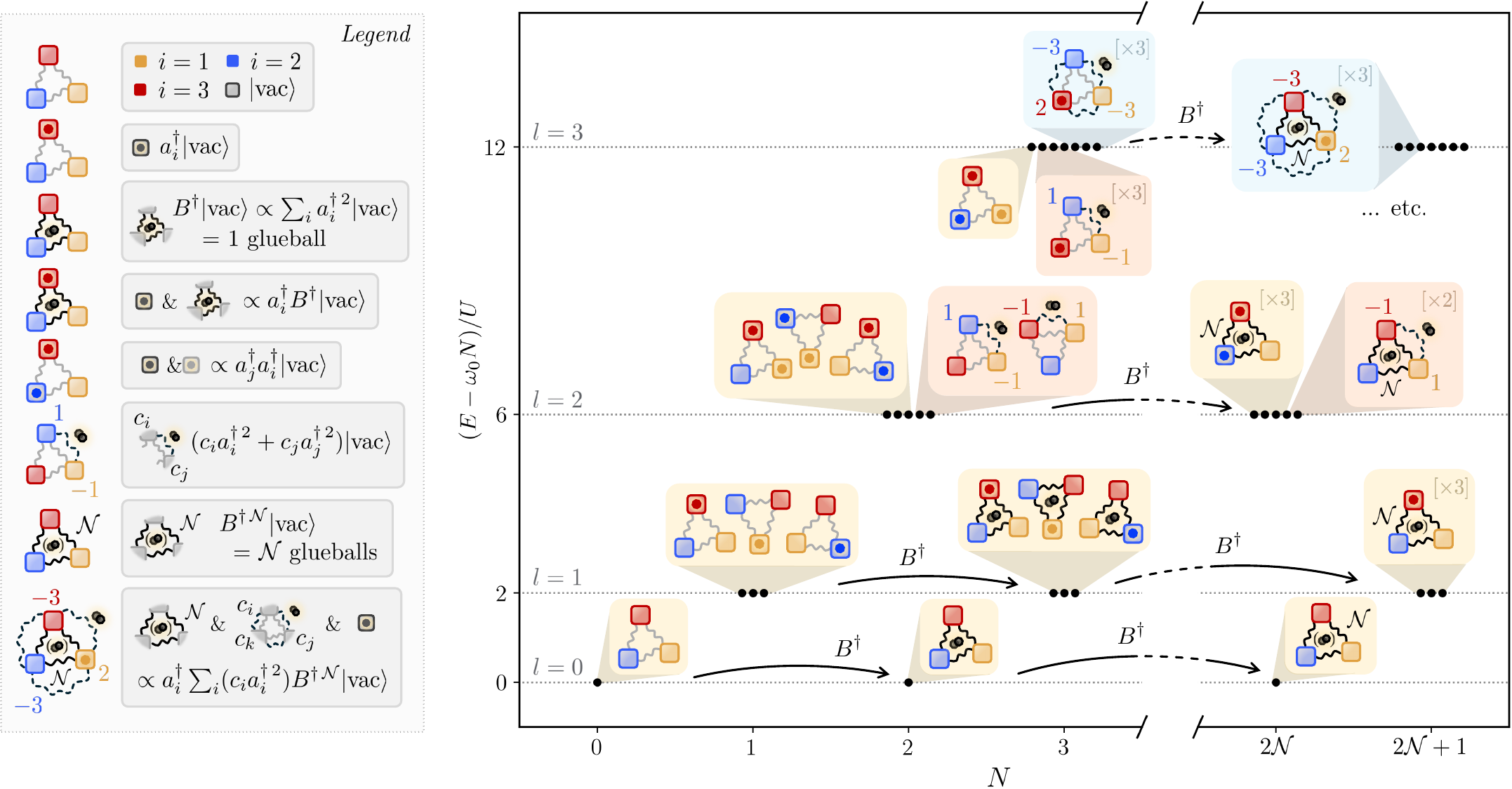}
    \caption{Energy spectrum of the cavity-SU(2) Hamiltonian for arbitrary excitation number, shown for angular-momentum sectors $l=0,1,2,3$.
\textit{Legend (left).}
Colored squares (red, blue, yellow) denote the three SU(2) gauge modes. Circles represent excitations: colored circles correspond to excitations localized in a single gauge mode of the same color, while black circles indicate excitations distributed across multiple modes.
A pair of black circles located at the center of the three modes (connected by a solid black wavy line) represents a glueball state, i.e. a composite two-excitation particle equally distributed among all gauge modes. A pair of black circles located outside the center (connected by dashed black wavy lines) represents two excitations unevenly shared between two or three modes; in these cases, the colored numbers indicate the relative amplitude of each gauge mode in the shared excitations.
When a glueball pair is enclosed in brackets and labeled by $\mathcal{N}$, this denotes a $\mathcal{N}$-glueball state, corresponding to $N=2\mathcal{N}$. Gray lines indicate inactive connections (no shared excitation pair). Diagrammatic elements are cumulative: multiple symbols represent products of the corresponding creation operators acting on the vacuum.
\textit{Spectrum (right).}
The plot displays eigenstate configurations in the local flavor basis for different values of $N$ and $l$. States are grouped into shaded boxes (yellow, orange, blue), which indicate the different families of states discussed in the text. 
For a fixed $l$, all configurations at excitation number $N$ can be reproduced at $N+2\mathcal{N}$ by acting on the states with the operator $(B^\dagger)^{ \mathcal{N}}$. Consequently, each $l$ sector exhibits a repetitive structure along the $N$ axis. 
}
    \label{fig:summary_maintext}
\end{figure*}

The introduction of the angular momentum operators $L_i$ provides a systematic way to label and construct all the eigenstates of $H_{\rm cav}$ using Eq.~\eqref{eq:general_eigenstate_exp}.
However, the choice of quantization axis singles out one flavor, here $i=3$, and from the general expression given in Eq.~\eqref{eq:general_eigenstate_exp}, the decomposition of the eigenstates in terms  of the individual gauge fields is not immediately obvious. 
In the following we summarize some additional relations and properties that allow us to express and also visualize the most relevant low-energy eigenstates in a flavor-symmetric manner.

We start with the identity 
\begin{align}\label{eq:Jsquare_BdagB}
    L^2 = N(N+1) - 6B^\dagger B,
\end{align}
which can be used to express the cavity Hamiltonian in terms of $N$ and the glueball ladder operators only. This relation implies that a state~$\ket{\phi}$ with $N$ bosons is an eigenstate of maximal angular momentum~$l=N$ if and only if it is a vacuum of glueballs, i.e., 
\begin{align}\label{eq:B_phi_0}
    B\ket{\phi} = 0.
\end{align}
We remark that $B^\dagger B$ is not the glueball number operator, since $B$ and $B^\dag$ are not bosonic operators. Instead, we obtain $B^\dagger B\ket{\mathcal{N}}_g=\mathcal{N}(2\mathcal{N}+1)/3 \ket{\mathcal{N}}_g$ (see Appendix~\ref{sec:supp:singlemode_algebraic}), such that Eq.~\eqref{eq:Jsquare_BdagB} is consistent with $L^2\ket{\mathcal{N}}_g=0$. 

Together with the useful commutation relation $[B,a_i^\dagger]=\sqrt{2/3} a_i$, Eq.~\eqref{eq:Jsquare_BdagB} and the property in Eq.~\eqref{eq:B_phi_0} can be used to identify the most relevant low-energy eigenstates with $l>0$, which are visualized in Fig.~\ref{fig:summary_maintext}.  For example, from Eq.~\eqref{eq:Jsquare_BdagB} it follows immediately that $a_i^\dagger\ket{\vac}$ is a $l=1$ eigenstate for all $i=1,2,3$, that $a_i^\dagger a_j^\dagger\ket{\vac}$ is a $l=2$ eigenstate for $i\neq j$, and that $a_1^\dagger a_2^\dagger a_3^\dagger\ket{\vac}$ is an eigenstate with $l=3$. 
We can repeat this construction by starting from a $\mathcal{N}$-glueball state $\sim (B^\dag)^\mathcal{N}|{\rm vac}\rangle$ instead of the vacuum. Since adding glueballs doesn't change the angular momentum quantum numbers, we obtain a whole family of eigenstates with $l=0,1,2,3$ and different total boson numbers. This family of states is indicated by the yellow shading in Fig.~\ref{fig:summary_maintext}. 

Similarly, we can identify another family of $l=2$ and $l=3$ states, given by $(a_i^{\dagger 2}-a_j^{\dagger 2})\ket{\vac}$ for $i\neq j$ and $a_i^\dagger(a_j^{\dagger 2}-a_{k}^{\dagger 2})\ket{\vac}$ for $i,j,k$ all different, respectively. These eigenstates are visualized by the orange boxes in Fig.~\ref{fig:summary_maintext}, and correspond to two bosons in an even superposition between two flavors, and zero or one boson in the last flavor. Again, an arbitrary number of glueballs can be added on top of these combinations without affecting the angular momentum quantum number. Finally, for $l=3$ we obtain a third family of states that contains states of the form $a_i^\dagger(2 a_i^{\dagger 2}-3a_j^{\dagger 2}-3a_k^{\dagger 2})\ket{\vac}$ for $i,j,k$ all different. They are represented in blue in Fig.~\ref{fig:summary_maintext} and can again contain an arbitrary amount of additional glueballs. 

While this construction becomes more tedious as the angular momentum quantum number increases, it is very useful to categorize and visualize the most relevant low-energy states in a manner that treats all flavors equally. This representation also provides a more intuitive understanding of transition matrix elements or other properties in terms of the original Fock states, rather than angular momentum quantum numbers. Note, however, that states obtained in this way are eigenstates of $H_{\rm cav}$ and linearly independent, but not necessarily orthogonal.

\subsection{Selection rules}\label{sec:selection_rules}

Since the operator $B$ removes glueballs, the property in Eq.~\eqref{eq:B_phi_0} implies that a $N$-boson state~$\ket{\psi}$ satisfying
\begin{equation}
    B^p\ket{\psi} = 0
\end{equation}
contains at most $p-1$ glueballs on top of maximal angular momentum states, such that it decomposes only on the subspaces $N\geq l>N-2p$. This property is useful to derive selection rules. Starting from a maximal angular momentum state $\ket{l,l,m}$ and applying the commutator  $[B^2,a_i^\dagger]=\sqrt{8/3}a_iB$, we find that $B^2 a_i^\dagger\ket{l,l,m}=0$. As such, the $N=l+1$ boson state $a_i^\dagger\ket{l,l,m}$ decomposes on the subspaces of angular momentum $l\pm 1$, and so does the state $a_i^\dagger\ket{N,l,m}$ due to the ladder property of $B^\dagger$.
Moreover, in the eigenbasis $\ket{N,l,m}$ corresponding to the choice of quantization axis $i=3$, we have $[a_3^\dagger,L_3]=0$, such that $a_3^\dagger$ preserves the quantum number $m$. This leads to the selection rule
\begin{align}\label{eq:selection_rule_a3}
    \bra{N',l',m'}a_3^\dagger\ket{N,l,m} \propto \delta_{N',N+1}\delta_{l',l\pm 1}\delta_{m',m}.
\end{align}
The corresponding matrix elements are derived in Appendix~\ref{sec:supp:singlemode_algebraic}.
The selection rules for $a_1^\dagger$ and $a_2^\dagger$ are obtained by expressing them in terms of the modes $a^\dagger_\pm$, which, satisfying $[L_3,a_\pm^\dagger]=\pm a_\pm^\dagger$, change the value of~$m$ by $\pm1$. We thus obtain
\begin{align}\label{eq:selection_rule_a12}
    \bra{N',l',m'}a_{i=1,2}^\dagger\ket{N,l,m} \propto \delta_{N',N+1}\delta_{l',l\pm 1}\delta_{m',m\pm1}.
\end{align}
These selection rules and the flavor symmetry of the glueball states imply a relation for the bosonic operators in the glueball subspace
\begin{align}
    \gluebra{\mathcal{N}}a_i^\dagger a_j\ket{\mathcal{N}}_g &=\frac{2\mathcal{N}}{3}\delta_{ij}, \\
    \gluebra{\mathcal{N}-1}a_i a_j\ket{\mathcal{N}}_g &=\delta_{ij} \sqrt{\frac{2}{3}}\gluebra{\mathcal{N}-1}B\ket{\mathcal{N}}_g,\label{eq:B_exp_glueball}
\end{align}
which we use to derive effective glueball Hamiltonians discussed below.

\section{Driven cavity}\label{sec:DrivenCavity}

While the single-mode SU(2) cavity model is an interacting system, its absolute ground state is the trivial vacuum state $|{\rm vac}\rangle$. To observe any nontrivial behavior, the system must be excited. This can be achieved, for example, by an external, time-dependent current, which would appear as an additional term $\sim J^{\rm ex}_{i,\mu}(t) A_i^\mu$ in the initial Lagrangian.  Within the single-mode approximation and assuming a monochromatic drive at a frequency~$\omega_{\rm d}$, this scenario can be modeled in terms of the driven SU(2) Hamiltonian
\begin{equation}
H_{\rm cav}(t) = \omega_0 N + \cavcoupling L^2 +  2\varepsilon\cos(\omega_{\rm d} t)(a_3 + a_3^\dagger),
\label{eq:Hlab_drive}
\end{equation}
where $\varepsilon$ is the strength of the external driving field.  Note that a more general driving term $\sim \sum_i \varepsilon_i (a_i+a_i^\dagger)$ can always be mapped to the form given in Eq.~\eqref{eq:Hlab_drive} through an appropriate rotation of the gauge flavors, $a_i\rightarrow R_{ij}a_j$, or, equivalently, by aligning the `quantization axis' along the direction of the external current.  

To proceed, we move into a rotating frame with respect to $\omega_{\rm d}$ by applying the unitary transformation $H_{\rm cav}\rightarrow \mathcal{U} H_{\rm cav}\mathcal{U}^\dag+ i\dot{\mathcal{U}} \mathcal{U}^\dag$, where $\mathcal{U}(t) =e^{i\omega_{\rm d} N t}$. Then, after making a rotating-wave approximation (RWA), which is valid under the condition~$\varepsilon\ll \omega_{\rm d}$, we obtain the time-independent Hamiltonian 
\begin{equation}
H_{\rm cav} = -\Delta N + \cavcoupling L^2 + \varepsilon(a_3 + a_3^\dagger),
\label{eq:Hrot_drive}
\end{equation}
where $\Delta=\omega_{\rm d}-\omega_0$ is the detuning of the drive from the bare cavity resonance. 

\subsection{Resonant excitation of a glueball condensate}
We first consider the case of a weak driving field with $\varepsilon, |\Delta| \ll \cavcoupling$, such that the glueball states $|\mathcal{N}\rangle_g$ can be resonantly excited. There is, however, no direct matrix element that couples the glueballs to the initial vacuum state, i.e., $\gluebra{\mathcal{N}} a_3^\dag+a_3 |{\rm vac}\rangle=0$, and this excitation process can only occur via virtual transitions to energetically higher states in the $l=1$ manifold. Using second-order perturbation theory and  the properties of the unperturbed eigenstates discussed in Sec.~\ref{sec:flavor_symm_states} above, we derive an effective Hamiltonian (see Appendix~\ref{app_PerturbationTheory})
\begin{equation}
H_{\rm eff} =  -\left(\Delta +\frac{\varepsilon^2}{3\cavcoupling}\right) N  -\frac{\varepsilon^2}{\sqrt{6}\cavcoupling}\,(B + B^\dagger),
\label{eq:Heff_drive}
\end{equation}
which describes the excitation dynamics within the $l=0$ glueball subspace. 

The Hamiltonian $H_{\rm eff}$ in Eq.~\eqref{eq:Heff_drive} is reminiscent of the Hamiltonian of an externally driven photonic mode in a conventional cavity QED setting, where $B$ and $B^\dagger$ would be replaced by bosonic creation and annihilation operators $a$ and $a^\dag$. In the photonic case, the cavity mode would be driven into a coherent state $|\alpha(t)\rangle$, with an amplitude $\alpha(t)\sim t$ that grows linearly in time when exact resonance conditions are met. In a particle picture, a coherent state $|\alpha\rangle$ corresponds to a condensate with a nonvanishing field expectation value, $\langle a\rangle\neq 0$. This analogy suggests to use a weakly driven SU(2) model to resonantly excite an equivalent glueball condensate. However, since the glueball ladder operators are not bosonic operators, 
the dynamics generated by $H_{\rm eff}$ is more complicated and a direct correspondence with a photonic coherent state fails. 

To investigate the weakly driven SU(2) cavity,  we remark that $B$ and $B^\dag$ are quadratic in the gauge-boson operators, such that $H_{\rm eff}$ can still be exactly diagonalized through a squeezing transformation~\cite{Gerry2005IntroductoryQuantumOptics}. Specifically, for $\Delta=-\varepsilon^2/(3\cavcoupling)$, the resulting evolution operator is given by 
\begin{equation}
e^{-i H_{\rm eff} t} = S_1(\xi_t)S_2(\xi_t)S_3(\xi_t),
\label{eq:psi_squeezing}
\end{equation}
where
\vspace{-5pt}
\begin{equation}
S_i(\xi)=\exp\!\left[\frac{1}{2}\big(\xi^\ast a_i^2-\xi\, a_i^{\dagger 2}\big)\right]
\label{eq:squeezing_parameter}
\end{equation}
is a single-mode squeezing operator and $\xi_t = -i \varepsilon^2 t/(3\cavcoupling)$.

From the perspective of conventional cavity QED, this result is significant for two reasons. First of all, even though $H_{\rm eff}$ describes a linear driving of glueball states, their composite, multi-photon nature results in an excitation dynamics that is very different from that of a driven photonic cavity. 
Secondly, the special structure of cavity SU(2) still permits an exact solution and, despite its interacting nature, the driven system evolves into three independently squeezed gauge bosons. For an initial state $|\psi\rangle(t=0)=|{\rm vac}\rangle$, this property allows us to reexpress the time-evolved state as $|\psi\rangle(t)=\sum_\mathcal{N} c_\mathcal{N}(t) |\mathcal{N}\rangle_g$, and derive exact expressions for the glueball amplitudes, $c_\mathcal{N}(t)= \gluebra{\mathcal{N}}e^{-i H_\eff t}\ket{\vac}$. We obtain 
\begin{align}\label{eq:driven_k_distribution}
    c_\mathcal{N}(t)=
    \frac{\sqrt{(2\mathcal{N}+1)!}[i\tanh(|\xi_t|)]^\mathcal{N}  }{2^\mathcal{N}\mathcal{N}!\cosh(|\xi_t|)^{3/2}}.
\end{align}
From this result, or by applying the squeezing transformation to the field operators~$a_i$, we can further derive the total glueball population,
\begin{align}\label{eq:mean_K}
    \langle \mathcal{N}\rangle(t) = \frac{3}{2}\sinh^2\left(\frac{\varepsilon^2 t}{3\cavcoupling}\right),
\end{align}
and the glueball field expectation value,
\begin{align}\label{eq:mean_B}
    \langle B\rangle(t) = i\frac{\sqrt{6}}{4}\sinh\left(\frac{2\varepsilon^2 t}{3\cavcoupling}\right).
\end{align}
Note that none of the individual modes acquires a non-vanishing field expectation value and $\langle a_i\rangle=0$. 

\begin{figure}
    \centering
    \includegraphics[width=
    \linewidth]{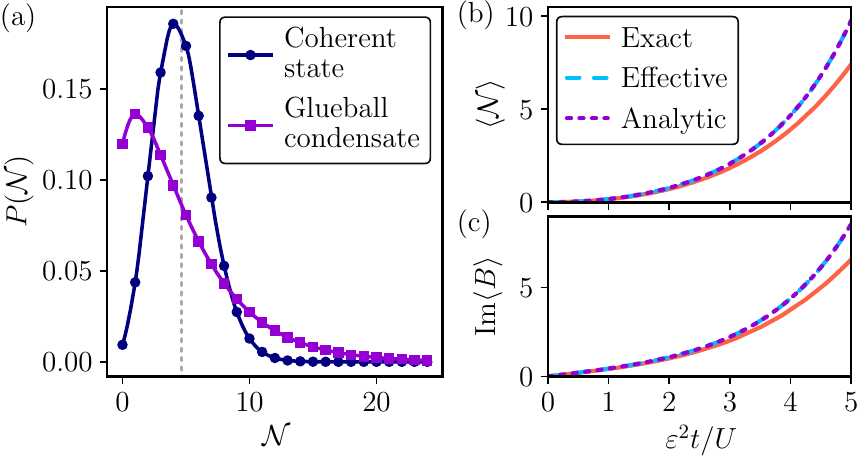}
    \caption{(a) Plot of the glueball number distribution $P(\mathcal{N})=|c_{\mathcal{N}}|^2$ obtained from Eq.~\eqref{eq:driven_k_distribution} at time $ t = 4\cavcoupling/\varepsilon^2$, corresponding to a squeezing parameter $|\xi_t|\simeq 1.33$. The resulting distribution (purple) is compared to a coherent-state distribution with the same average glueball number (blue). The vertical dashed line indicates the average glueball number $\langle \mathcal{N} \rangle\simeq 4.7$.
    (b) and (c) Time evolution of the average glueball number $\langle \mathcal{N} \rangle$ and amplitude $\mathrm{Im}\langle B\rangle$, as obtained from simulations of the  full driven Hamiltonian in Eq.~\eqref{eq:Hrot_drive} (solid lines) with $\varepsilon/U=0.2$, the effective model $H_{\rm eff}$ in Eq.~\eqref{eq:Heff_drive} (dashed lines), and the analytic results derived in Eq.~\eqref{eq:mean_K} and Eq.~\eqref{eq:mean_B} (dotted lines). For the numerical simulations, we have assumed a detuning of $\Delta=-\varepsilon^2/3U$ and restricted the Hilbert space to states with $m=0$.}
    \label{fig:drive_distribution}
\end{figure}

In Fig.~\ref{fig:drive_distribution}(a) we plot the glueball populations $P(\mathcal{N})=|c_\mathcal{N}|^2$ of a weakly driven SU(2) cavity at a time $t=4U/\varepsilon^2$ and compare it---as a reference---to the photon populations $|\langle \mathcal{N} |\alpha\rangle|^2$ of a coherent state with the same average occupation number. In particular, we see that the glueball distribution has a much larger width, $\Delta \mathcal{N}=\sqrt{\langle \mathcal{N}^2\rangle-\langle \mathcal{N}\rangle^2}$, when compared to that of a coherent state. Using the moments of the photon number in a squeezed state, we obtain
\begin{align}
    \frac{\Delta \mathcal{N}}{\langle \mathcal{N}\rangle}=\sqrt{\frac{2}{3}} \coth\left(\frac{\varepsilon^2 t}{3\cavcoupling}\right) \underset{t\gg \cavcoupling/\varepsilon^2}{\simeq}\sqrt{\frac{2}{3}}.
\end{align}
This shows that in a driven glueball condensate, fluctuations of the populations always remain large compared to its average value.

We emphasize that the analytic results for the glueball condensate derived above are valid in the weak-driving regime, and the exact dynamics will eventually deviate from the exponential growth predicted in Eq.~\eqref{eq:mean_K}. This occurs when the matrix element $\varepsilon\bra{2\mathcal{N}+1,1,0}a_3^\dagger\ket{\mathcal{N}}_g$ is of the order of the gap, $2\cavcoupling$, restricting the validity of the effective model to glueball numbers $\mathcal{N}_{\max}\lesssim \cavcoupling^2/\varepsilon^2$. In Fig.~\ref{fig:drive_distribution}(b,c) we show the dynamics of $H_{\rm eff}$ and compare it to the predictions of the full Hamiltonian in Eq.~\eqref{eq:Hrot_drive}. We see that, over the range shown in Fig.~\ref{fig:drive_distribution}(b,c), the full numerical evolution closely follows the analytic prediction for both $\langle \mathcal{N}\rangle$ and $\mathrm{Im}\langle B\rangle$. This comparison reaches dimensionless times $\varepsilon^2 t/U \simeq 5$, corresponding to an average of $\langle\mathcal{N}\rangle\approx 10$ glueballs, which can be accessed numerically by exploiting the selection rule that the driven dynamics starting from the vacuum remains restricted to the $m=0$ subspace.

\subsection{Rabi oscillations and revivals in cavity SU(2)}
Within the weak-driving regime, $\varepsilon \ll U$, but allowing for arbitrary detunings $\Delta$, we can select the resonance condition 
\begin{equation}
\Delta N = \cavcoupling l(l+1)  
\end{equation}
to resonantly excite other higher lying states. Here we illustrate this process for the case $l=1$ and $\Delta=2\cavcoupling$. 
Under this condition, any state with $l=0$ and $\mathcal{N}$ glueballs, $|2\mathcal{N},0,0\rangle$, is resonant with the state $|2\mathcal{N}+1,1,0\rangle$ and, according to Eq.~\eqref{eq:selection_rule_a3}, there is a nonvanishing matrix element of the drive term that couples these pairs of states. At the same time, transitions to the $l=2$ manifold as well as transitions to other states in the  $l=1$ and $l=0$ subspaces with fewer or more glueballs are energetically forbidden. Therefore, under these conditions, we obtain a family of coupled two-level subspaces,
\begin{align}\label{eq:Rabi_Ham_Delta=2}
    H_{\rm cav} \simeq \sum_{\mathcal{N}=0}^\infty \frac{\Omega_\mathcal{N}}{2}\left(\ket{2\mathcal{N}+1,1,0}\bra{2\mathcal{N},0,0} + {\rm H.c.}\right),
\end{align}
where the Rabi frequency
\begin{align}\label{eq:rabifreq}
    \Omega_\mathcal{N} = 2\sqrt{\frac{2\mathcal{N}+3}{3}}\varepsilon
\end{align}
is obtained from the matrix elements of $a_3^\dagger$ given in Appendix~\ref{sec:supp:singlemode_algebraic}.

\begin{figure}
    \centering
    \includegraphics[width=\linewidth]{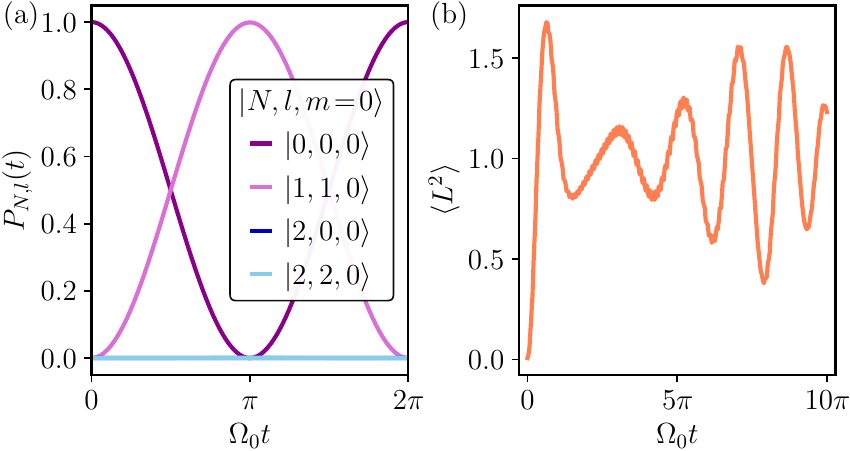}
    \caption{(a) Rabi oscillations induced by a resonant drive with $\varepsilon/U= 0.05$ and detuning $\Delta=2\cavcoupling$, showing coherent population transfer between the states $\ket{0,0,0}$ and $\ket{1,1,0}$. Populations $P_{N,l}=|\bra{N,l,0}|\psi\rangle|^2$ of higher-energy states remain vanishingly small, confirming that the dynamics is effectively restricted to an isolated two-level subspace with Rabi frequency $\Omega_0 = 2\varepsilon$.
    (b) Collapse and revival of the angular momentum expectation value $\langle L^2\rangle$ under the same driving conditions as in (a), but for a system initialized in a glueball condensate. This initial state is prepared by evolving the vacuum under a weak drive with $\Delta=-\varepsilon^2/(3\cavcoupling)$ for a time $t_{\mathrm{prep}}=3\cavcoupling/\varepsilon^2$, corresponding to a squeezing parameter $|\xi_t|=1$. Both plots are obtained from exact numerical simulations of the full Hamiltonian in Eq.~\eqref{eq:Hrot_drive}. 
    }
    \label{fig:drive_Rabi_11}
\end{figure}

Starting from the vacuum state, $H_{\rm cav}$ couples only a single pair of states, resulting in Rabi oscillations between $|{\rm vac}\rangle$ and $|1,1,0\rangle$, as illustrated in Fig.~\ref{fig:drive_Rabi_11}(a). By stopping the drive after a time $t=\pi/\Omega_0$, this scheme provides a way to deterministically prepare a single boson state. Alternatively, the system can be initialized in a glueball condensate, before the resonant excitation to the $l=1$ manifold is applied. In this case, $H_{\rm cav}$ couples multiple pairs of states, but with incommensurate Rabi-frequencies $\Omega_\mathcal{N}\sim \sqrt{2\mathcal{N}+3}$. The resulting evolution then exhibits a rapid initial collapse of the oscillation, followed by a partial revival at a later time. This feature is reminiscent of the phenomenon of collapse and revival in cavity QED~\cite{haroche2013ExploringQuantumAtoms}, where the incommensurate frequencies arise from the coupling of photons to matter. Here the intrinsic properties of the gauge bosons already give rise to a similar effect.

We remark that also energetically higher-lying states can be accessed via an external drive by choosing an appropriate resonance condition. For example, to prepare the state $\ket{N,l,m}=\ket{3,1,0}$, one can drive the system with a detuning~$\Delta\simeq 2\cavcoupling/3$, which makes it resonant with the vacuum state. However, according to the selection rules of $a_3^\dagger$, the two states will only be coupled via a three-boson process. Such higher-order driving schemes typically require larger driving amplitudes and a more refined analysis of Stark shifts and the leakage to non-resonant states~\cite{Ruyten1989,Sah2024,Xia2025,Schirk2025,Huang2026}. In general, any state $\ket{N,l,m=0}$ can be prepared by an $N$-photon process with $\Delta\simeq \cavcoupling l(l+1)/N$.

\section{Waveguide cavity SU(2)}\label{sec:waveguide_su2}

\begin{figure}
    \centering
    \includegraphics[width=\linewidth]{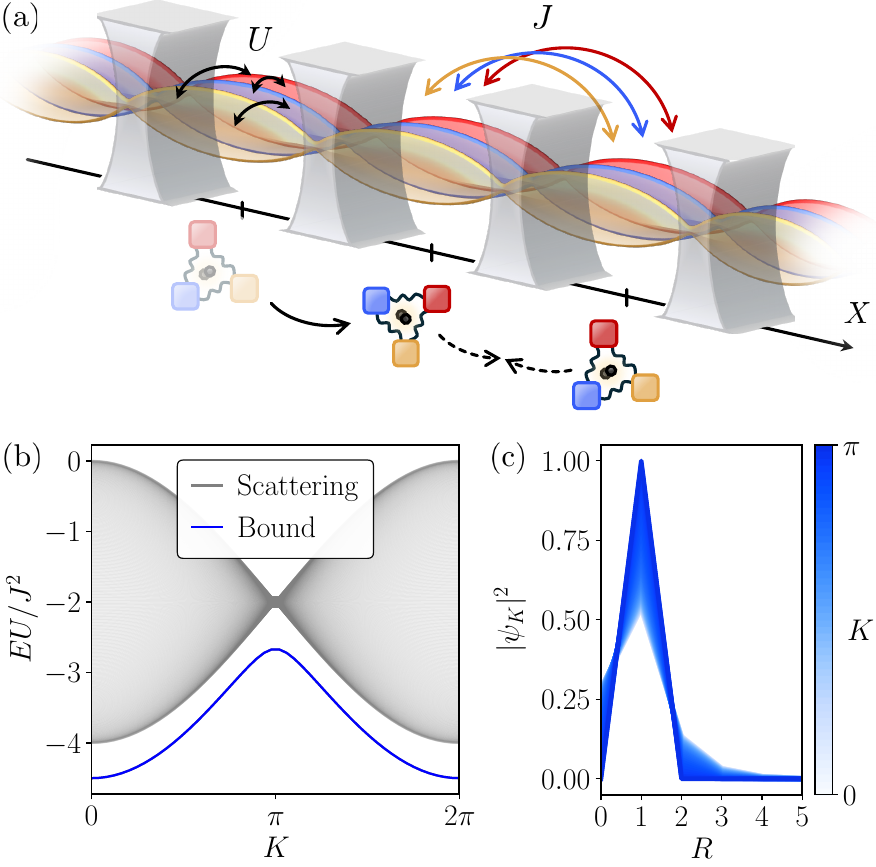}
    \caption{(a) Sketch of an SU(2) cavity array. Each site hosts three interacting gauge-boson modes with local interaction strength $\cavcoupling$, while neighboring cavities are coupled by single-boson hopping with amplitude $J$. In the regime of strong interactions, $\cavcoupling\gg J$, the low-energy excitations are local glueballs, represented schematically by two photons symmetrically distributed among the three modes. Virtual hopping processes generate effective glueball hopping of order $J^2/\cavcoupling$, illustrated by the solid arrows and a faded glueball diagram, as well as nearest-neighbor density-density interactions, illustrated by the dashed arrows. (b) Two-glueball (four-boson) spectrum in the limit $J\ll\cavcoupling$. The gray band corresponds to scattering states of two independent glueballs, while the isolated blue branch below the continuum corresponds to a two-glueball bound state. (c)  Probability distribution $|\psi_K(R)|^2$ of the bound state in (b), where $R=X-Y$ is the separation between the two glueballs and $K$ is the center-of-mass momentum. The distribution is localized at short distances and peaks around $R=1$, indicating that the bound state preferentially occupies neighboring cavities.}
    \label{fig:bound_states}
\end{figure}

Within the single-mode approximation, the spatial distribution of the gauge fields is fixed by the selected mode function $\vec u_0(\vec x)$, which determines the bare energy of the confined gauge bosons as well as their mode-averaged mutual interactions. Motion in space can be reintroduced in this framework by adding the coupling to more and more orthogonal modes $\vec u_{n\neq0}(\vec x)$, which become closer spaced in energy and more relevant as the quantization volume $V\rightarrow \infty$ increases. Here we consider a different scenario, illustrated in Fig.~\ref{fig:bound_states}, where extended space is represented by an array of coupled cavities, each of finite volume $V$. In this setting, the single-mode approximation is still justified within each cavity, but the gauge bosons can tunnel between neighboring sites through a residual evanescent overlap between the respective mode functions.  Similar to electrons in a tight-binding lattice potential, this tunneling leads to a delocalization of the bosons and the formation of a finite propagation band.  This scenario is analogous to coupled-cavity arrays studied for conventional photons~\cite{Yariv1999,GobanNatComm2014,LodahlRMP2015,liu_quantum_2017,ChangRevModPhys2018,Ferreira2021,JouannyNatComm2025}, where the tunneling amplitude determines the effective propagation speed and introduces an adjustable kinetic energy term that competes with local interactions.

\subsection{SU(2) cavity array}\label{sec:Array}
To extend our framework from a single cavity to a 1D array, we envision a chain of $M\gg1$ adjacent cavities, labeled by the discrete index $X$. The cavities are arranged along the $x$-direction and separated by sharp, weakly transparent potential walls. In this case, the former localized modes $\vec u_n(\vec x)$ that diagonalize the noninteracting Hamiltonian $H_0$ are replaced by extended Bloch waves, 
\begin{equation}
\vec u_n(\vec x)  \rightarrow \vec f_{n,k}(\vec x),
\end{equation} 
where $n$ is the band index and $k\in [-\pi/d,\pi/d]$ the quasi momentum. The Bloch waves are periodic in the lattice spacing $d$, i.e.,  $\vec f_{n,k}(\vec x+d\vec e_x)= e^{ikd} \vec f_{n,k}(\vec x)$. In the limit where the barriers are high, the eigenfrequencies $\omega_{n,k}$ separate into a set of narrow, energetically isolated bands, and we can restrict the expansion of the gauge fields to modes $\vec f_k\equiv \vec f_{0,k}$ within a single band with index $n=0$ and frequencies $ \omega_{0,k}\approx \omega_0$.

To make the connection to the single-cavity Hamiltonian $H_{\rm cav}$, it is convenient to use a basis of Wannier functions, which are defined as the Fourier transform of the Bloch mode with respect to the quasi-momentum,
\begin{align}\label{eq:supp:Wannier_mode_def}
    \vec w_{X}(\vec x) &= \frac{1}{\sqrt{M}}\sum_k e^{-ikd X}\vec f_{k}(\vec x). 
\end{align}
In the limit of well-separated cavities, these Wannier functions reduce to the fully localized modes $\vec u_0$ within each cavity, i.e., $\vec w_{X}(\vec x+dX\vec e_x)\simeq \vec u_0(\vec x)$. Therefore, for bosons occupying the same mode, we obtain similar interaction terms as in the single-cavity case. However, the Wannier modes are no longer eigenfunctions of the free Hamiltonian. Apart from a less relevant common frequency shift, this induces a nonvanishing matrix element between neighboring sites, $J= -\langle \vec w_{X}| H_0 |\vec w_{X+1}\rangle \neq 0$. When expressed in terms of the bosonic operators $a_{i,X}$ for the Wannier modes, this correction translates into the hopping Hamiltonian
\begin{equation}
    H_{\mathrm{hop}}
    = -J \sum_{X}\sum_{i=1}^3
    \left(a_{i,X}^{\dagger}\, a_{i,X+1} + \mathrm{H.c.}\right).
    \label{eq:hop}
\end{equation}
In addition, even in the limit where the Wannier modes are strongly localized, the Coulomb term in $\mathcal{H}_{\rm int}$ gives rise to long-range interactions that induce nonlinear processes between gauge bosons in different cavity sites. Therefore, in the limit of well-separated but not fully isolated cavities, the single-band Hamiltonian for the SU(2) array is of the form (see Appendix~\ref{app:mode_expansion_array})
\begin{equation}\label{eq:Harray}
    \begin{aligned}
    H_{\mathrm{array}} = ~&\sum_X \left( \omega_0 N_X + \cavcoupling L_X^2\right)+  H_{\mathrm{hop}}   \\
    & + \frac{1}{2}\sum_{X\neq X'} \tilde U_C(X,X')L_{X,i}L_{X',i},
\end{aligned}
\end{equation}
where $N_X$, $L_{X,i}$ and $L_X^2$ are the number operators and angular momentum operators for the individual lattice sites. The second line of Eq.~\eqref{eq:Harray} represents the long-range part of the Coulomb interaction with 
\begin{align}
    \tilde U_C(X,X^\prime) & = \frac{g^2}{4\pi}\int d^3xd^3y\, \frac{|\vec w_X(\vec x)|^2|\vec w_{X^\prime}(\vec y)|^2}{|\vec x-\vec y-d(X-X^\prime)\vec e_x|}\nonumber \\
    &  \approx \frac{g^2}{4\pi d} \frac{1}{|X-X^\prime|}.
\end{align}

\subsection{Strong-coupling regime and effective glueball dynamics}\label{sec:Heff_array}

Compared to the single-cavity model, the Hamiltonian $H_{\rm array}$ describes a competition between local, repulsive interactions between the gauge bosons and the kinetic energy term $H_{\rm hop}$, which tends to delocalize them across the whole lattice. This competition gives rise to interesting many-body effects, which, however, are no longer exactly solvable and beyond the scope of the current analysis.  To understand, nevertheless, some of the essential few-body processes that can occur in this SU(2) cavity array, we focus on the regime $\cavcoupling \gg J$, where the strong on-site interactions dominate over the hopping. In the limit $J\rightarrow 0$, and for a given total boson number, the energetically lowest states are formed by glueballs arbitrarily distributed across the lattice. Within this subspace, the hopping term $H_{\rm hop}$ acts as a perturbation that induces motion and effective interactions between these composite objects.

Degenerate perturbation theory up to second order in $J/\cavcoupling$ yields a compact effective Hamiltonian governing glueball motion and effective glueball-glueball interactions in the strong-coupling regime,
\begin{equation}\label{eq:Heff}
    \begin{aligned}
    H_\eff 
    = -\frac{J^2}{\cavcoupling} \sum_{X}& \left[\mathcal{N}_X + \frac{2}{3} \mathcal{N}_X \mathcal{N}_{X+1} \right. \\
    &+ \left.\frac{1}{2}\left(B_{X+1}^\dagger B_{X} + B_{X+1} B_{X}^\dagger \right) \right],
\end{aligned}
\end{equation}
where $\mathcal{N}_X$ is the glueball number operator in cavity~$X$.
The first term is a global energy shift, or effective chemical potential. The second term is a density-density or cross-Kerr interaction between neighboring cavities. The last term represents a glueball hopping term, written in terms of the glueball ladder operators~$B_X$ and $B_X^\dagger$. 
Note that, since glueballs are $l=0$ states, the inter-cavity Coulomb interaction vanishes identically in the glueball subspace. For simplicity, higher-order modifications of the effective hopping strength from the long-range part of the Coulomb interaction in the excited subspace have been neglected in Eq.~\eqref{eq:Heff}. See Appendix~\ref{sec:array_Heff} for more details on the derivation of $H_{\rm eff}$. 

We remark that lattice Hamiltonians similar to the model in Eq.~\eqref{eq:Harray} have been analyzed in the context of spin-1 bosons in optical lattices~\cite{DemlerZhou2002, Imambekov2003, Rizzi2005}. In these models, however, additional spin-independent on-site interactions prevent a clustering of glueballs on the same site, which is not the case in the SU(2) cavity system.  

\subsection{Two-glueball bound states}

Starting from the effective Hamiltonian in Eq.~\eqref{eq:Heff}, we first consider a two-glueball system in an infinitely long chain, $M\rightarrow \infty$. The two-glueball state can be decomposed as
\begin{align}\label{eq:two_glueball_state}
    \ket{\Psi}=\frac{1}{\sqrt{2}}\sum_{X,Y} c_{X,Y} \Psi(X,Y)B^\dagger_X B^\dagger_Y\ket{\vac},
\end{align}
where $c_{X,Y} =\sqrt{2/C_{\mathcal{N}=2}}$ for $X=Y$ and $c_{X,Y}=1$ otherwise. Using this convention, $\Psi(X,Y)$ is normalized to 
\begin{align}
    \sum_{X,Y}|\Psi(X,Y)|^2 =1 
\end{align}
and $|\Psi(X,Y)|^2$ can be interpreted as the probability for finding two glueballs at sites $X$ and $Y$.

By translation symmetry, the center-of-mass quasi-momentum~$K$ is a conserved quantum number, such that the eigenstates can be found using the ansatz~$\Psi_K(X,Y)=e^{-iK(X+Y)/2}\psi_K(X-Y)/\sqrt{M}$. This separation maps the Schr\"odinger equation into a single particle equation for the wavefunction of the relative coordinate $R=X-Y\geq 0$, which we can restrict to positive values due to the symmetry of the wavefunction. The resulting eigenvalue equation reads
\begin{equation}\label{eq:BoundStateEq}
\begin{gathered}
    E_K\psi_K(R) = -\dfrac{J^2}{\cavcoupling}\Bigg[ \left(2+\dfrac{2\delta_{R1}}{3}\right)\psi_K(R)  \qquad \\
    +\cos\left(\frac{K}{2}\right)\Big(t_R\,\psi_K(R{+}1)+t_{R-1}\,\psi_K(R{-}1)\Big)\Bigg],
\end{gathered}
\end{equation}
where $t_R = 1 + (\sqrt{C_{\mathcal{N}=2}}-1)\delta_{R0}$. This result has been derived from Eq.~\eqref{eq:Heff} and Eq.~\eqref{eq:two_glueball_state} by making use of the properties of the glueball operators derived in Sec.~\ref{sec:Spectrum_of_SU2}.

Solving Eq.~\eqref{eq:BoundStateEq} numerically on a finite lattice, we obtain the two-glueball spectrum shown in Fig.~\ref{fig:bound_states}(b). For each value of $K$, we obtain a continuum of scattering states, which represent two distant glueballs propagating through the lattice independently. In addition, there is one state which is energetically separated and lies below the scattering band. The corresponding wavefunction for the relative coordinate, $\psi_K(R)$, is plotted in Fig.~\ref{fig:bound_states}(c). It exhibits a sharp maximum around $R=1$, indicating a bound state of two glueballs (four gauge bosons) that  propagate through the lattice jointly and are on average one site apart. This is the composite-particle counterpart of the two-particle bound states known from Bose-Hubbard-type models for atoms in optical lattices~\cite{Winkler2006, ValientePetrosyan2008} and interacting photons in regular cavity arrays~\cite{Ma2019,Morvan2022_short,Wang2020}.

\subsection{Few-body bound states}

\begin{figure}
    \centering
\includegraphics[width=\columnwidth]{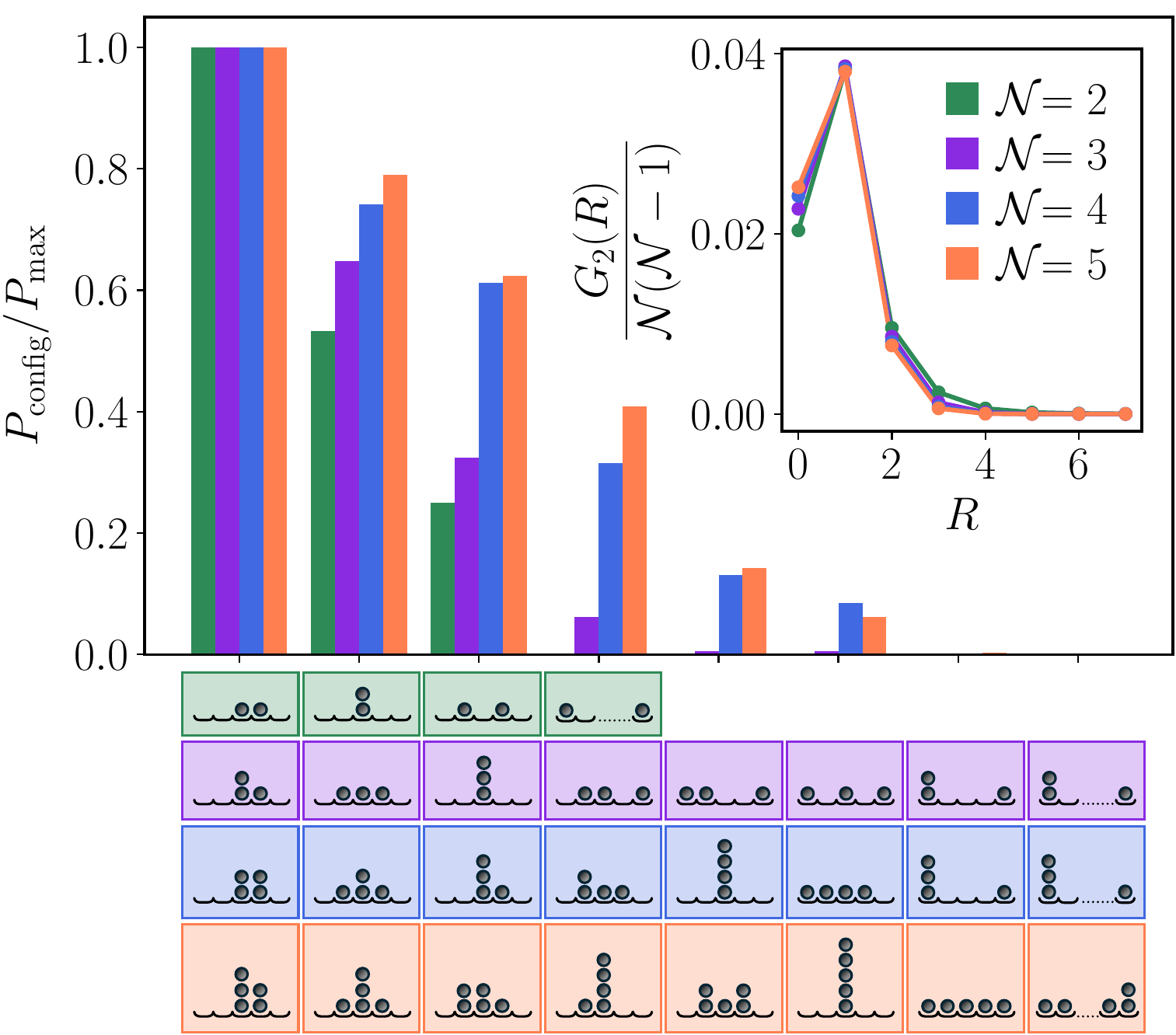}
    \caption{Configuration probabilities and pair correlations in the multi-glueball sector.
    \textit{Main panel:} normalized probabilities $P_\text{config}/P_\text{max}$ for representative occupation-number configurations in the $\mathcal{N}$-glueball sectors ($\mathcal{N}=2,3,4,5$). The configurations shown along the horizontal axis are illustrative examples, chosen to compare compact clustered states with increasingly separated, low-weight configurations (green, purple, blue and orange frames indicate the $\mathcal{N}=2,3,4,5$ configurations). Clustered configurations in which glueballs occupy adjacent sites carry the largest weight, while configurations containing one or more spatial gaps are strongly suppressed, indicating a large energetic cost for separating glueballs from the cluster.
    \textit{Inset:} Two-particle correlator $G_2(R)=\langle :\!\mathcal{N}_0\mathcal{N}_R\!:\rangle$, summed over both directions on the ring for $0<R<M/2$, normalized by the total number of ordered particle pairs $\mathcal{N}(\mathcal{N}-1)$. The correlator is peaked at $R=1$ and decreases rapidly with separation.}
    \label{fig:manybody_boundstates}
\end{figure}

In this section we present qualitative evidence for the formation of many-body bound states in the multi-glueball sector. While a semi-analytic characterization of the full spectrum was possible for the two-glueball case, such an analysis becomes intractable for larger glueball numbers $\mathcal{N}$. Instead, we perform numerical simulations to evaluate the lowest-energy eigenstate $|\Psi\rangle$ of $H_{\rm eff}$ within a sector of fixed glueball number, which is conserved by $H_{\rm eff}$, on a ring of $M=14$ cavities with periodic boundary conditions and for $\mathcal{N}=2,3,4,5$. The ring size was chosen such that the configurations shown in Fig.~\ref{fig:manybody_boundstates} include separations up to $R=M/2=7$.
The $\mathcal{N}=2$ sector serves as a reference to connect with the two-glueball analysis above. This state can be expanded in a basis $|\mathcal{N}_1,\mathcal{N}_2,\dots, \mathcal{N}_M\rangle$ specifying the numbers of glueballs in each lattice site, from which we can extract the relative probabilities of different glueball configurations. Since it has vanishing center-of-mass momentum, $K=0$, all configurations related by a translation of the ring carry the same weight, and only one representative of each is shown. In this context, binding is reflected by a strong suppression of configurations in which one or more glueballs are spatially separated from the rest of the particles.

Figure~\ref{fig:manybody_boundstates} shows the normalized probabilities for selected representative configurations in the $\mathcal{N}$-glueball sectors. For all cases, the dominant contributions are compact configurations, in which glueballs occupy adjacent lattice sites. Any configuration containing a gap--that is, one or more glueballs separated from the main cluster by at least one empty site--is strongly suppressed. This suppression persists even when separated subgroups remain clustered among themselves, indicating an energetic penalty originating from the spatial separation itself, rather than from local rearrangements within each subgroup.

Within the compact sector, configurations with distributed occupation across neighboring sites are favored over maximal on-site occupation. In particular, the largest probability weight is obtained when the glueballs are approximately evenly distributed across two adjacent sites. This behavior is consistent with the two-glueball bound state analyzed previously, where the attractive cross-Kerr interaction favors the nearest-neighbor configuration over the on-site one. The resulting many-body bound state is therefore not an on-site condensate but a spatially extended composite object with internal structure.

To quantify spatial correlations, we introduce the normal-ordered density-density correlation function
\begin{equation}
G_2(R)=\langle :\!\mathcal{N}_0 \mathcal{N}_R\!:\rangle =  \langle (\mathcal{N}_0-\delta_{0R}) \mathcal{N}_R\rangle.
\end{equation}
Upon normalization by the total number of ordered pairs, $\mathcal{N}(\mathcal{N}-1)$, the quantity $G_2(R)/[\mathcal{N}(\mathcal{N}-1)]$, summed over both directions on the ring, equals $|\psi_K(R)|^2/M$ in the two-glueball case and is constant for $0<R<M/2$ for uncorrelated glueballs.

The inset of Fig.~\ref{fig:manybody_boundstates} shows the normalized correlator $G_2(R)/[\mathcal{N}(\mathcal{N}-1)]$ for $\mathcal{N}=2,3,4,5$. The correlations are strongly peaked at short distances and decay rapidly with increasing separation. As the glueball number increases, the qualitative structure of both the configuration probabilities and the correlation functions remains unchanged. Taken together, these observations provide consistent evidence for the emergence of spatially compact many-body bound states in the multi-glueball sector.

\section{Analog simulation of cavity SU(2)}\label{sec:analog_simulation}
Apart from its conceptual appeal and structural simplicity, an important benefit of introducing a cavity framework for SU(2) is that it makes the theory accessible for near-term analog quantum simulation. In view of the bosonic nature of its fundamental constituents and the Kerr-type interactions, superconducting quantum circuits provide a natural platform to realize such schemes. Previously, various superconducting circuit configurations have already been proposed for simulating lattice gauge theories and quantum link models with finite-dimensional representations of the gauge fields~\cite{Marcos2013, Marcos2014, Mezzacapo2015, Brennen2016, Zhang2023,  Busnaina2025}. In our current cavity formalism, the continuous gauge fields can be directly represented by the harmonic excitations of microwave $LC$ resonators, while their mutual  interactions can be engineered with the help of additional Josephson nonlinearities. In the following we outline a basic circuit design that is well-suited for realizing both the model of a single SU(2) cavity as well as that of a cavity array.

\subsection{Circuit design and Hamiltonian engineering}

To implement the single-cavity Hamiltonian in Eq.~\eqref{eq:H_singleMode_expanded}, we seek a superconducting circuit that generates cross-Kerr interactions of the form $a_i^\dagger a_ia_j^\dagger a_j$ and two-photon hopping processes $a_i^{\dagger\,2}a_j^2$, while suppressing other undesired nonlinearities, like self-Kerr terms of the form $a_i^{\dagger\,2}a_i^2$. The main challenge is therefore to engineer strong nonlinear interactions between three microwave modes without introducing large local anharmonicities.

\begin{figure}
    \centering
    \includegraphics[width=\linewidth]{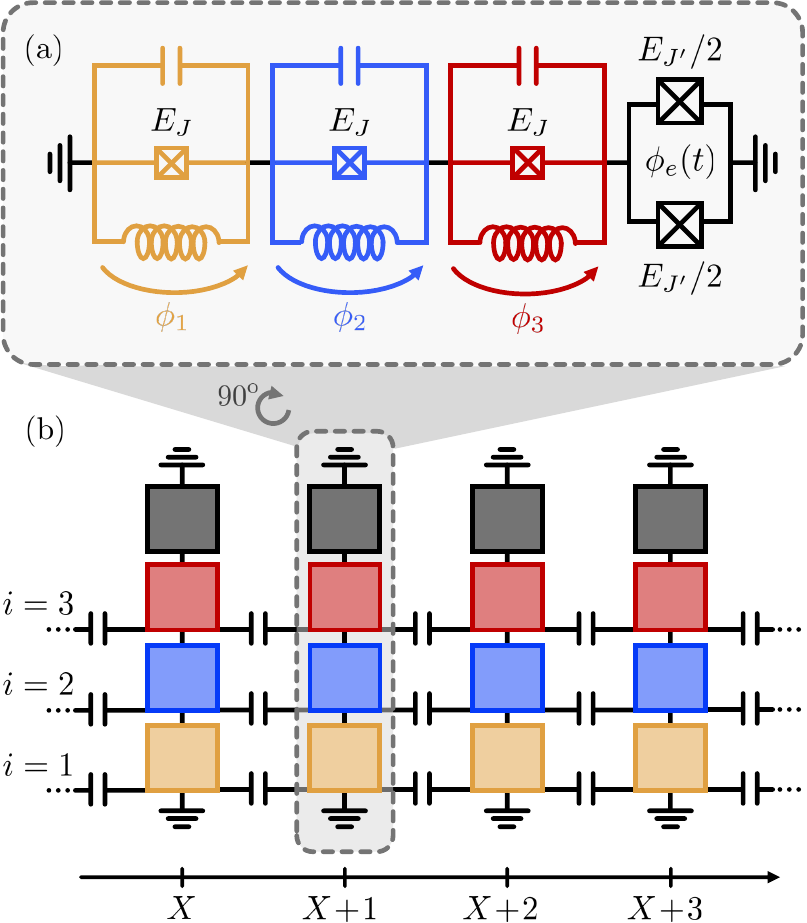}
    \caption{(a) Sketch of a superconducting circuit layout for realizing the SU(2)-cavity Hamiltonian, $H_{\rm cav}$. The circuit consists of three nonlinear $LC$ resonators, representing the gauge bosons of different flavor. These subcircuits share a current path and are shunted by a SQUID loop. The circuit parameters and the flux $\phi_e(t)$ threading the SQUID loop are engineered to produce the Hamiltonian in Eq.~\eqref{eq:H_singleMode_expanded}. See additional details about the SQUID loop in Appendix~\ref{app:circuit}. (b) The circuit units in (a) can be arranged into a capacitively coupled array to realize the waveguide SU(2) model in Eq.~\eqref{eq:Harray} (for $\tilde U_C\simeq 0$).   
    }
    \label{fig:implementation}
\end{figure}

\subsubsection{Cavity modes} 
To achieve this goal, we propose the superconducting circuit shown in Fig.~\ref{fig:implementation}{(a)}. The circuit consists of three lumped-element $LC$ resonators, each representing one flavor of gauge bosons. Each resonator contains a Josephson junction to induce local nonlinearities and they are coupled in series to another auxiliary Josephson  element (drawn in black). 
The latter is controlled by an external magnetic flux $\phi_e(t)$ (in units of the flux quantum and accompanied by a flux $\phi_e^\prime(t)=-\phi_e(t)/2$ through the loop to ground) to engineer effective interactions across the chain.  The Hamiltonian for this circuit reads (see Appendix~\ref{app:circuit})
\begin{equation}
\begin{aligned}\label{eq:Hcircuit1}
    H_{\rm circ}  =\,& \sum_i\left(4E_C^iq_i^2+\dfrac{E_L^i}{2}\phi_i^2\right)- \sum_i E_{J} \cos{(\phi_i)} \\
    &  \qquad - E_J' \cos\left(\frac{\phi_e(t)}{2}\right)\cos\left(\sum_i \phi_i\right)\\
    \equiv \,  & H_{\rm har} + H_{\rm nl}(t),
\end{aligned}
\end{equation}
Here, the $\phi_i$ and $q_i$ are the dimensionless canonical flux and charge variables for the subcircuits, and they obey $[\phi_i,q_i]=i$.  For each resonator,  $E^i_{L}$,  $E_{C}^i$, and $E_{J}$ denote the inductive, capacitive and Josephson energies, respectively, and $E_J^\prime$ is the effective Josephson energy of the auxiliary coupling element.

We consider the external flux $\phi_{e}(t)=2[\pi+\theta_0+\theta_t(t)]$ with $|\theta_t(t)|\ll 1$,  and we assume $E^i_C\ll E^i_L,E_J$. Under these conditions, we can expand the Josephson potentials in Eq.~\eqref{eq:Hcircuit1} in powers of $\phi_i$ and $\theta_t(t)$ and keep only the most relevant low-order contributions. After this expansion, we group all time-independent terms that are quadratic in $q_i$ and $\phi_i=\phi_\text{ZPF}^i(a_i+a_i^\dag)$ into the harmonic Hamiltonian $H_{\rm har}$, which can be written in terms of the bosonic annihilation (creation) operators $a_i$ ($a_i^\dag$) for each resonator mode as
\begin{equation}
H_{\rm har}=\sum_i \omega_i a_i^\dag a_i. 
\end{equation}
In the following we assume that the ratio $E_C^i/\tilde E_L^i$, where $\tilde E^i_L=E_L^i+E_J-E_J^\prime \cos(\theta_0)$, is the same for all three resonators, while the products $E_C^i\tilde E_L^i$ are dissimilar. This assumption implies that all modes have the same zero-point fluctuations  $\phi_\text{ZPF}^i = (2E_C^i/\tilde E_L^i)^{1/4}\equiv \phi_\text{ZPF}$, but distinct resonance frequencies, $\omega_i = (8E_C^i \tilde E_L^i)^{1/2}$. For superconducting microwave circuits, these frequencies are typically in the range of $3-10$ GHz, and  we can select frequencies $\omega_i$ with non-equidistant spacings $\delta_{ij}=\omega_i-\omega_j\sim 1-2$ GHz within this range.

As we discuss below, the choice of  distinct resonator frequencies is beneficial for engineering the targeted nonlinear interactions, but it also breaks the symmetry between the three modes. For quantum simulation of a SU(2) cavity, we thus proceed with a unitary transformation, $\tilde H_{\rm circ} =  e^{iH_0t}H_{\rm circ}e^{-iH_0t}$, into a rotating frame with respect to $H_0 = \sum_i \omega_{r,i} a_i^\dag a_i$. In this frame and choosing $\omega_{r,i}= \omega_i-\omega_0 - 2U$, we obtain 
\begin{equation}
\tilde H_{\rm har}=  \omega_0 N + 2U  \sum_i n_i, 
\end{equation}
which represents the noninteracting part of the cavity SU(2) Hamiltonian in Eq.~\eqref{eq:H_singleMode_expanded}.

\subsubsection{Cross-Kerr interactions}
The nonlinear contribution $H_{\rm nl}$ in Eq.~\eqref{eq:Hcircuit1} contains all the remaining terms of the expansion. We keep those terms up to fourth order in the fluxes $\phi_i$ and set $\theta_t(t)=0$ for now. In this case and moving to the rotating frame introduced above, terms $\sim \phi_i$ and $\sim \phi_i\phi_j\phi_k$, containing an odd number of bosonic operators, oscillate rapidly, at least with the frequencies $\omega_{r,i}$, and cancel out on average. Similarly, out of the remaining quadratic and fourth-order terms, all processes that do not conserve the occupation number of each mode individually oscillate at least with the mode detunings $\delta_{ij}$. For large mode spacings, as assumed above, also these contributions are similarly suppressed. Therefore, we can approximate $\tilde H_{\rm nl}=\tilde H_{\rm res}+\tilde H_{\rm non-res}(t)\approx \tilde H_{\rm res}$, where the remaining resonant contribution obeys $[\tilde H_{\rm res},a_i^\dag a_i]=0$ and is given by
\begin{equation}
\begin{aligned}
 \tilde H_{\rm res} = & - \frac{\phi_\text{ZPF}^4}{2} \left[ E_J-3 E^\prime_J \cos(\theta_0) \right]  \sum_i a_i^\dag a_i \\
 & - \frac{\phi_\text{ZPF}^4}{4} \left[ E_J-E^\prime_J \cos(\theta_0) \right]  \sum_i a_i^{\dag 2} a_i^2 \\
& + \frac{\phi_\text{ZPF}^4}{2} E^\prime_J \cos(\theta_0) \sum_{i\neq  j}a_i^\dag a_i a_j^\dag a_j.
\end{aligned}
\end{equation}
This result shows that by choosing  $E_J=E_J^\prime \cos(\theta_0)$, we can eliminate the  self-Kerr interactions $\sim a_i^{\dag 2} a_i^2$, while the small frequency corrections $\sim a_i^\dag a_i$ can be absorbed into a redefinition of the individual $\omega_i$. As a result, we are left with  
\begin{equation}\label{eq:HCrossKerr}
\tilde H_{\rm res} \simeq  U \sum_{i\neq j}a_i^\dag a_i a_j^\dag a_j,
\end{equation}
where $U= E_J \phi_\text{ZPF}^4/2$. This is the cross-Kerr term that appears in the cavity SU(2) model in Eq.~\eqref{eq:H_singleMode_expanded}. Note that in the parameter regime of interest, the interaction strength  $U\sim E_C E_J/E_L \ll \omega_i,|\delta_{ij}|$ is consistent with the omission of other non-resonant fourth-order terms.

\subsubsection{Two-photon hopping}
While the choice of well-separated resonator frequencies is convenient to suppress many of the unwanted three- and four-photon processes, this choice also eliminates  the desired two-photon hopping processes $\sim a_i^{\dagger\,2}a_j^2$.  To reintroduce those contributions in a resonant and controlled manner, we now consider a weak, oscillating component of the external flux, 
\begin{equation}
 \theta_t(t)=\sum_{i<j}A\cos(\omega_{ij}^d t) ,\qquad   \omega_{ij}^d=2(\omega_i-\omega_j).
\end{equation}
For a sufficiently weak driving amplitude $A\ll 1$, this additional off-resonant drive does not affect any of the approximations made above, except that it makes the set of terms $\sim a_i^{\dagger\,2}a_j^2$ in $\tilde H_{\rm non-res}(t)$ time-independent. Therefore, we can approximate $\tilde H_{\rm non-res}(t)\approx \tilde H_{\rm 2hop} $, where
\begin{equation}\label{eq:H2PhotonHopp}
\tilde H_{\rm 2hop} =  -A \frac{\phi_\text{ZPF}^4}{8} E^\prime_J \sin(\theta_0)  \sum_{i\neq j}a_i^{\dag 2} a_j^2.
\end{equation}
Then, by adjusting the driving amplitude to $A  = 8U/[\phi_\text{ZPF}^4 E_J^\prime \sin(\theta_0)]$, this term contributes the remaining two-boson hopping terms in Eq.~\eqref{eq:H_singleMode_expanded}.

\subsection{Parameters and Scalability}
In summary, we have shown that up to fourth order in the flux variables and a sufficiently weak flux driving, the Hamiltonian for the three coupled circuits is well-approximated by 
\begin{equation}
\tilde H_{\rm circ}\simeq H_{\rm cav} + H_{\rm osc}(t),
\end{equation} 
where $H_{\rm cav}$ is the SU(2) cavity Hamiltonian of interest. The residual correction terms in $H_{\rm osc}(t)$ oscillate at frequencies $\omega_{i}$ and $\delta_{ij}$. For the weakly nonlinear, inductively shunted resonators considered here, $U=E_JE_C/E_L$ and typical values are $U/2\pi\sim 1-10$ MHz~\cite{Grimm2020},
which is much smaller than the assumed frequency separations between the modes. This condition justifies the omission of all non-resonant Kerr interactions, while the dynamics associated with $H_{\rm cav}$ still remains fast compared to typical lifetimes in superconducting circuits, $T_1\gtrsim 100\,\mu$s~\cite{Blais2021, Krantz2019}, with millisecond lifetimes demonstrated for transmon qubits~\cite{Bland2025} and single-photon internal quality factors above $10^6$ for superconducting resonators~\cite{Reagor2013,Bruckmoser2026}.
Note that in order to match the cross-Kerr terms in Eq.~\eqref{eq:HCrossKerr} and the two-photon processes in Eq.~\eqref{eq:H2PhotonHopp}, we
require 
\begin{equation}
A= \frac{4}{\tan(\theta_0)}.
\end{equation}
In view of the assumption $A\ll1$, the current circuit design requires a rather large asymmetry $E_J^\prime \gg E_J$. This potential drawback can be mitigated by adding additional junctions or fluxes for the modulation without affecting the overall Hamiltonian engineering strategy. Note that various techniques for cancelling self-Kerr interactions~\cite{Frattini2017,Sivak2019} or for realizing cross-Kerr couplings between microwave modes~\cite{Holland2015} are already routinely implemented in circuit QED experiments.

A key advantage of the proposed circuit architecture is its modularity. Each three-resonator unit, together with its nonlinear coupler, constitutes a self-contained ``SU(2) cavity'' building block. A one-dimensional chain, as considered in Sec.~\ref{sec:Array}, can then be realized by arranging such units linearly and introducing additional coupling capacitances between neighboring modes of the same flavor, as illustrated schematically in Fig.~\ref{fig:implementation}(b). Under a RWA, this capacitive coupling implements a tunneling Hamiltonian, as given in Eq.~\eqref{eq:hop}. In this way, the $M$-site SU(2) waveguide Hamiltonian in Eq.~\eqref{eq:Harray} (with $\tilde U_C=0$) can be realized with a $4M$-element superconducting circuit architecture.  The interactions remain analog and locally engineered, providing a hardware-efficient route toward larger systems, and thus allowing us to probe the many-body features of non-Abelian cavity gauge theories.

\section{Conclusion}\label{sec:conclusion}

In summary, we have presented a cavity framework for pure SU(2) gauge theories in which the gauge fields are represented by a minimal set of spatial modes. In its simplest realization, the resulting cavity Hamiltonian, $H_{\rm cav}$,  describes three independent modes coupled through cross-Kerr interactions and two-photon hopping processes that directly reflect the underlying structure of the gauge group. We have analyzed the resulting energy spectrum and identified composite glueball states as the lowest-energy excitations within each particle-number sector. Despite the presence of repulsive boson-boson interactions, these glueballs can be resonantly driven by an external field into a glueball condensate with characteristic pair- or squeezing correlations. Moreover, when the system is extended to arrays of tunnel-coupled cavities, we predict the formation of bound states of glueballs, which may otherwise propagate freely through the lattice. These examples already demonstrate that the physics of SU(2) cavity systems can differ profoundly from the behavior expected in conventional cavity-QED settings with nonlinear interactions.

The cavity formalism introduced in this work was both motivated and designed to retain a close analogy with conventional cavity QED. By working in the Coulomb gauge and expanding the gauge fields in terms of discrete eigenmodes of the noninteracting theory, we established a direct correspondence between cavity QED and its non-Abelian counterpart. This approach reveals the similarities as well as the differences between those theories and shows how the additional direct and Coulomb-mediated couplings between the gauge fields are affected by the choice of modes in the expansion. It is therefore complementary to the more common framework of lattice gauge theories, in which the noninteracting and interacting parts of the theory are discretized on an equal footing at a fundamental length scale, and where this analogy with cavity QED is obscured. 

Another important feature of the cavity-mode expansion is that it renders SU(2) gauge theories with continuous degrees of freedom accessible to near-term quantum-simulation platforms. For example, the gauge bosons can be mapped onto equivalent photonic cavity modes in the optical or microwave regime. To demonstrate the feasibility of this approach, we proposed a concrete realization of SU(2) cavities and SU(2) cavity arrays based on superconducting quantum circuits. However, similar quantum-simulation schemes may also be envisioned in trapped-ion platforms \cite{Calajo2024, John2026, Saner2025} and ultracold-atom systems \cite{GonzalezCuadra2022, HalimehSpinExchange2024}, where a comparable degree of control over bosonic degrees of freedom can be achieved experimentally.

\section*{Data Availability}
Numerical data and code to recreate the figures are available at \href{https://github.com/lucia-valor/cavity-models-of-non-Abelian-gauge-theories}{github.com/lucia-valor}.

\begin{acknowledgments}
We thank Nora Brambilla, Zohreh Davoudi, Daniel González Cuadra, Fabian Grusdt and Federico Roy for valuable feedback and discussions. We acknowledge financial support from the German Federal
Ministry of Education and Research via the funding program under contract number 13N16188 (MUNIQC-SC) and under contract number 13N15680 (GeQCoS). This research is part of the Munich Quantum Valley, which is supported by the Bavarian state government with funds from the Hightech Agenda Bayern Plus. LV is partially supported by the Rafael del Pino Foundation. 
\end{acknowledgments}

\appendix

\section{The longitudinal field component}\label{sec:Elong_derivation}

In this section, we derive the SU(2) Yang-Mills Hamiltonian in the Coulomb gauge, following the general analysis of~\cite{Abers1973GaugeTheories,Abers197313YangMillsField}.
We decompose the electric field in its longitudinal and transverse components,  $\vec E_{i}=\vec E_{i}^\perp+\vec E_{i}^\parallel$, where $\vec\nabla\cdot\vec E_{i}^\perp=0$, and $\vec\nabla\times\vec E_{i}^\parallel=0$. We then introduce a scalar potential $\phi_i$ such that $\vec E_{i}^\parallel = -\vec\nabla\phi_i$. The Gauss law in Eq.~\eqref{eq:Elong_Gausslaw} leads to a set of coupled linear equations for the potentials
\begin{align}\label{eq:Gauss_potential_eq}
    \left(\delta_{ik}\Delta - g\varepsilon_{ijk}\vec A_j\cdot\vec\nabla\right)\phi_k = - g\rho_i,
\end{align}
where $\rho_i=\varepsilon_{ijk}\vec A_j\cdot\vec E_k^\perp$.
In the perturbative regime, this equation can be solved using Green functions, by inverting the differential operator perturbatively in~$g$. We denote by 
\begin{align}\label{eq:M_op}
    V_{ik}=\varepsilon_{ijk}\vec A_j \cdot\vec\nabla
\end{align}
the perturbation and by 
\begin{align}
    G_{ik}(\vec x,\vec y)=-\frac{\delta_{ik}}{4\pi|\vec x-\vec y|}
\end{align}
the Green function of the Laplacian in three dimensions. Then, by using vector notations for the flavor indices, $\vec{\Phi}=(\phi_1,\phi_2,\phi_3)$, etc., we obtain the formal solution
\begin{align}\label{eq:formal_inv_pot}
   \vec{\Phi} = - g   (\mathbf{G} + g\mathbf{GVG}+g^2\mathbf{GVGVG}+...) \vec{\rho},
\end{align}
where the operator products are convolutions. This enables a systematic expansion of the longitudinal fields in terms of the transverse ones in the perturbative regime. At leading order, we get the usual electrostatic potential induced by a source charge,
\begin{align}
    \phi_i(\vec x) =  g\int d^3y \; \dfrac{\rho_i(\vec y)}{4\pi|\vec x-\vec y|}+\mathcal{O}(g^2).
\end{align}
The longitudinal electric field energy is then obtained by an integration by parts
\begin{align}
    \int d^3x~ \vec E_{i\parallel}\cdot\vec E_{i\parallel}&=-\int d^3 x~\phi_i \Delta\phi_i \\
    &=g^2 \int d^3 xd^3y~\frac{\rho_i(\vec x)\rho_i(\vec y)}{4\pi|\vec x-\vec y|}+\mathcal{O}(g^3).
\end{align}
Note that, in Eq.~\eqref{eq:M_op}, $\mathbf{V}$ is linear in the gauge fields. As such, higher-order corrections also correspond to terms with products of more fields.

Note that the Coulomb gauge condition does not fix the gauge completely. For sufficiently large fields, there exist gauge copies satisfying $\vec\nabla \cdot \vec A_i=0$~\cite{Gribov1978QuantizationNonAbelianGauge}, associated with zero-modes of the Faddeev–Popov operator $M_{ik}=\delta_{ik}\Delta-g\varepsilon_{ijk}\vec A_j\cdot\vec\nabla$~\cite{greensite2020IntroductionConfinementProblem} appearing in Eq.~\eqref{eq:Gauss_potential_eq}. A complete treatment restricts the configuration space to the fundamental modular region inside the Gribov region $\Omega$ where $M_{ij}> 0$~\cite{Zwanziger1998, vanBaal2000,FeuchterReinhardt2004}, both in the path-integral and in the Hamiltonian formulation. The perturbative expansion in $g$ used here is an expansion around $A = 0$, deep inside $\Omega$, and the effects of the boundary of the fundamental domain on low-lying states are non-perturbative in $g$~\cite{vanBaalKoller1987} and consistently neglected in an expansion in powers of $g$. In the single-mode model the field amplitude scales as 
$|A_i| \sim \sqrt{n_i}$, so the expansion is controlled for boson numbers $n\ll 1/g^2$ [see Eq.~\eqref{eq:a9}].

The above condition also ensures the validity of the weak-coupling assumption, which permits the restriction of the Coulomb interaction to its leading order. More quantitatively, the next order in Eq.~\eqref{eq:formal_inv_pot} is subdominant if
\begin{align}\label{eq:supp:Coulomb_small_condition}
    g\left|\int d^3y \frac{\varepsilon_{ijk}(\vec x-\vec y)\cdot \vec A_j(\vec x)}{4\pi |\vec x-\vec y|^3}\rho_k(\vec y)\right|\ll \left|\rho_i( \vec x)\right|.
\end{align}
Within the single-mode approximation, see Eq.~\eqref{eq:field_restriction_single_mode}, we can estimate the validity of this condition by assuming the scaling $\vec A_i \sim 1/\sqrt{\omega_0 V}$ and $\vec E_i^\perp \sim \sqrt{\omega_0/ V}$, where $V=\ell_c^3$ is the cavity mode volume. Up to numerical prefactors, we then obtain the condition 
\begin{equation}\label{eq:a9}
\sqrt{\frac{g^2}{\omega_0 \ell_c}}\ll 1
\end{equation}
for the validity of the weak-coupling expansion. This constraint is consistent with the assumption $U_C/\omega_0\ll 1$, i.e., that the resulting Coulomb interaction between the modes is small compared to their bare energies.

\section{Mode expansion}\label{sec:mode_expansion}

In this appendix, we outline the derivation of the interaction matrix elements in the mode expansion in Eq.~\eqref{eq:Hint_general_exp}. We write the interaction Hamiltonian as 
\begin{equation}
H_{\rm int}= H_3 + H_4 + H_C,
\end{equation}
where the individual terms represent the contributions from the three-mode and four-mode couplings and the Coulomb interaction, respectively. 

\subsection{Single-mode approximation}
We first look at the single-mode approximation in Eq.~\eqref{eq:field_restriction_single_mode}, where each gauge field is restricted to a single mode with the same mode function $\vec u(\vec x)$, where we have omitted the index $0$ for brevity. 

\subsubsection{Three-boson term}
The Hamiltonian $H_3$ contains three field operators and reads 
\begin{equation}
H_3= - g  \int d^3 x  \,\varepsilon_{ijk} \left(\partial_a A^b_{i}\right)A_j^a A_k^b.
\end{equation}
When expanding the kernel in terms of annihilation and creation operators, we obtain, for example, a term of the form 
\begin{equation}
\frac{\varepsilon_{ijk}}{2}  \vec u\cdot \vec \nabla(\vec u\cdot \vec u)   a_i a_j a_k =0.
\end{equation}
When summing over the Levi-Civita tensor, it adds up to zero since all the operators commute and the prefactor, which is determined by the same mode function, is also identical. This argument applies for all other combinations of annihilation and creation operators as well and $H_3=0$ within the single-mode approximation.

\subsubsection{Four-boson term}
To evaluate the contribution from the  four-field coupling, 
\begin{equation}
H_4= \frac{g^2}{4} \int d^3 x  \,\varepsilon_{ijk}\varepsilon_{ilm} \left(\vec A_{j} \cdot \vec A_{l}\right)  \left(\vec A_{k}  \cdot \vec A_{m}\right),
\end{equation}
we make use of the relation $\varepsilon_{ijk}\varepsilon_{ilm}=\delta_{jl}\delta_{km}-\delta_{jm}\delta_{kl}$ and obtain
\begin{equation}
H_4= \frac{g^2}{16\omega_0^2} \int d^3 x  \,\sum_{j\neq k} \left(X_{jj}X_{kk}-X_{jk}X_{jk} \right).
\end{equation}
Here, we have introduced the short-hand notation
\begin{equation}
X_{rs}= (\vec u\cdot \vec u ) a_r a_s + |\vec u|^2 \left( a_r^\dag a_s + a_s^\dag a_r\right) +(\vec u^*\cdot \vec u^* ) a^\dag_r a^\dag_s. 
\end{equation}
By expanding these products, we can collect all contributions with the same number of annihilation (creation) operators. The contribution $\sim a_j^{2}a_k^{2}$ vanishes, since all terms commute. The same is true for contributions of the form $(\vec u\cdot \vec u ) |\vec u|^2 a_j^\dag a_j a_k^2$, where all terms in the product cancel. The remaining terms are
\begin{align}
&\left(X_{jj}X_{kk}-X_{jk}X_{jk} \right)=\\
&=|\vec u|^2 |\vec u|^2 \left[ (2a_k^\dag a_k +1)(2a_j^\dag a_j +1) -(a_k^\dag a_j +a_j^\dag a_k)^2 \right] \nonumber\\
&+|(\vec u\cdot \vec u )|^2 \left[ a_k^{\dag2} a_j^2 + a_j^{\dag2} a_k^2- a_j a_j^\dag a_k a_k^\dag - a_j^\dag a_j  a^\dag_k a_k \right]. \nonumber
\end{align}
After bringing all terms into a normal-ordered form, we finally obtain 
\begin{align}
&\left(X_{jj}X_{kk}-X_{jk}X_{jk} \right)=\left(|\vec u|^2 |\vec u|^2 -|(\vec u\cdot \vec u )|^2\right) \nonumber\\
&\qquad \times\left(2 n_j n_k +n_k+n_j- a_k^{\dag2} a_j^2 - a_j^{\dag2} a_k^2\right),
\end{align}
which leads to the cross-Kerr interaction in $H_{\rm cav}$ with $U_4$ defined in Eq.~\eqref{eq:U4}. 

Alternatively, we can make use of the definition of the angular momentum operators $L_i = - i \varepsilon_{ijk} a_j^\dag a_k$ introduced in Sec.~\ref{sec:Spectrum_of_SU2} and write 
\begin{align}
 \varepsilon_{ijk}A_{j}^aA_k^b &= \frac{i}{2\omega_0}\left[ (u^a)^* u^b- u^a (u^b)^* \right] L_i
 \label{eq:supp:epsilonAE}.
\end{align}
In this case the relevant expression,
\begin{align}
\left(\varepsilon_{ijk}A_{j}^a A_{k}^b\right)^2 &=- \frac{L^2}{4\omega_0^2} \left[ (u^a)^* u^b- u^a (u^b)^* \right]^2 \nonumber\\
&=\frac{L^2}{2\omega_0^2} (|\vec u|^2 |\vec u|^2 -|(\vec u\cdot \vec u )|^2),
\end{align}
follows more directly.

\subsubsection{Coulomb term}
To evaluate the contribution from the Coulomb term, 
\begin{equation}
H_C= \frac{g^2}{8\pi} \int d^3 x d^3y \,\frac{\rho_i(\vec x) \rho_i(\vec y)}{|\vec x-\vec y|},
\end{equation}
we make again use of the angular momentum operators defined in Eq.~\eqref{eq:angular_momentum_def} and write the charge operators as
\begin{align}
    \rho_i(\vec x)=\varepsilon_{ijk}\vec A_j(\vec x)\cdot \vec E_k(\vec x) &=-|\vec u(\vec x)|^2 L_i. 
\end{align}
From this result, we immediately arrive at the Coulomb contribution to $H_{\rm cav}$ with $U_C$ defined in Eq.~\eqref{eq:UC}.

\subsection{Cavity array}\label{app:mode_expansion_array}
For the case of a cavity array considered in Sec.~\ref{sec:Array}, we expand the gauge fields in terms of Wannier modes, 
\begin{equation}
\vec A_i(\vec x)\simeq \sum_X \frac{1}{\sqrt{2\omega_0}} \left[ \vec w_X(\vec x) a_{i,X} + \vec w_X^*(\vec x) a_{i,X}^\dag\right].
\end{equation}
Since in the tight-binding limit the Wannier function $\vec w_X(\vec x)\approx \vec u_0(\vec x-dX\vec e_x)$ is strongly localized at site $X$, we can approximate 
\begin{equation}
\int d^3x  \, f(\vec x) \vec w_X(\vec x) \cdot\vec w_Y^*(\vec x)\approx \delta_{X,Y} \int_V d^3x \, f(\vec x) |\vec u_0(\vec x)|^2,
\end{equation}
etc. This implies that in $H_3$ and $H_4$ only local contributions, which are identical to the single-mode case, survive. 

Under the same approximation, we can write the charge density operator in the Coulomb term as  
\begin{align}
    \rho_i(\vec x)\simeq  \sum_X |\vec u_0(\vec x- dX  \vec e_x )|^2 L_{X,i}. 
\end{align}
This expansion gives rise to the additional long-range Coulomb interaction terms in Eq.~\eqref{eq:Harray}.

\subsection{General matrix elements}

When each of the gauge fields is expanded in terms of multiple modes, there are additional nonvanishing matrix elements, leading to the general expressions in Eq.~\eqref{eq:Hint_general_exp}. For completeness, here we summarize the corresponding interaction matrix elements. Note that, even after normal ordering, these coefficients are not unique. Since the sums in Eq.~\eqref{eq:Hint_general_exp} run over all indices and creation (annihilation) operators commute among themselves, only the total coefficient of each distinct operator product is fixed, for example, $G^{(0,2)}_{i,nm}+G^{(0,2)}_{i,mn}$ for $a_{i,n}a_{i,m}$. The expressions below are therefore one of several equivalent choices.

\begin{widetext}
\begin{align}
G_{\substack{i\\nm}}^{(0,2)}={}&
\frac{g^2}{4\sqrt{\omega_n\omega_m}}
\sum_p\frac{1}{\omega_p}\int d^3x\,
\Big[
|\vec u_p|^2(\vec u_n\!\cdot\!\vec u_m)
-(\vec u_p\!\cdot\!\vec u_n)
(\vec u_p^*\!\cdot\!\vec u_m)
\Big]
\nonumber\\
&+\frac{g^2}{16\pi}\sum_p
\int\frac{ d^3x\, d^3y}{|\vec x-\vec y|}
\frac{(\omega_p-\omega_n)(\omega_p+\omega_m)}
{\omega_p\sqrt{\omega_n\omega_m}}
\times \big[\vec u_n(\vec x)\!\cdot\!\vec u_p(\vec x)\big]
\big[\vec u_p^*(\vec y)\!\cdot\!\vec u_m(\vec y)\big],
\label{eq:Gfinal-02}
\end{align}

\begin{align}
G_{\substack{i\\nm}}^{(1,1)}={}&
\frac{g^2}{4\sqrt{\omega_n\omega_m}}
\sum_p\frac{1}{\omega_p}\int d^3x\,
\Big[
2|\vec u_p|^2(\vec u_n^*\!\cdot\!\vec u_m)
-u_p^a u_p^{b*}
\big(u_n^{a*}u_m^b+u_m^a u_n^{b*}\big)
\Big]
\nonumber\\
&+\frac{g^2}{16\pi}\sum_p
\int\frac{ d^3x\, d^3y}{|\vec x-\vec y|}
\frac{1}{\omega_p\sqrt{\omega_n\omega_m}}
\Big[
(\omega_m-\omega_p)(\omega_n-\omega_p)
\times
\big[\vec u_p(\vec x)\!\cdot\!\vec u_m(\vec x)\big]
\big[\vec u_p^*(\vec y)\!\cdot\!\vec u_n^*(\vec y)\big]
\nonumber\\[-6pt]
&\hspace{5.5cm}+(\omega_n+\omega_p)(\omega_p+\omega_m)
\big[\vec u_n^*(\vec x)\!\cdot\!\vec u_p(\vec x)\big]
\big[\vec u_p^*(\vec y)\!\cdot\!\vec u_m(\vec y)\big]
\Big],
\label{eq:Gfinal-11}
\end{align}

\begin{equation}
G_{\substack{ijk\\nmp}}^{(0,3)}=
-\frac{g\varepsilon_{ijk}}{\sqrt{8\omega_n\omega_m\omega_p}}
\int d^3x\,
(\partial_a u_n^b)u_m^a u_p^b,
\label{eq:G03}
\end{equation}

\begin{equation}
G_{\substack{ijk\\nmp}}^{(1,2)}=
-\frac{g\varepsilon_{ijk}}{\sqrt{8\omega_n\omega_m\omega_p}}
\int d^3x\,\Big[
(\partial_a u_n^{b*})u_m^a u_p^b-(\partial_a u_m^b)u_n^{a*}u_p^b
+(\partial_a u_m^b)u_p^a u_n^{b*}
\Big],
\label{eq:G12}
\end{equation}

\begin{align}
G_{\substack{ijkl\\nmpq}}^{(0,4)}={}&
\frac{g^2\varepsilon_{\alpha ij}\varepsilon_{\alpha kl}}
{32\sqrt{\omega_n\omega_m\omega_p\omega_q}}
\int d^3x\,
\Big[
(\vec u_n\!\cdot\!\vec u_p)(\vec u_m\!\cdot\!\vec u_q)
-(\vec u_n\!\cdot\!\vec u_q)(\vec u_m\!\cdot\!\vec u_p)
\Big]
\nonumber\\[2pt]
&-\frac{g^2\varepsilon_{\alpha ij}\varepsilon_{\alpha kl}}{128\pi}
\int\frac{d^3x\,d^3y}{|\vec x-\vec y|}
\frac{(\omega_m-\omega_n)(\omega_q-\omega_p)}
{\sqrt{\omega_n\omega_m\omega_p\omega_q}}
\times
\big[\vec u_n(\vec x)\!\cdot\!\vec u_m(\vec x)\big]
\big[\vec u_p(\vec y)\!\cdot\!\vec u_q(\vec y)\big],
\label{eq:Gfinal-04}
\end{align}

\begin{align}
G_{\substack{ijkl\\nmpq}}^{(1,3)}={}&
\frac{g^2\varepsilon_{\alpha ij}\varepsilon_{\alpha kl}}
{8\sqrt{\omega_n\omega_m\omega_p\omega_q}}
\int d^3x\,
\Big[
(\vec u_n^*\!\cdot\!\vec u_p)(\vec u_m\!\cdot\!\vec u_q)
-(\vec u_n^*\!\cdot\!\vec u_q)(\vec u_m\!\cdot\!\vec u_p)
\Big]
\nonumber\\[2pt]
&-\frac{g^2\varepsilon_{\alpha ij}\varepsilon_{\alpha kl}}{32\pi}
\int\frac{d^3x\,d^3y}{|\vec x-\vec y|}
\frac{(\omega_n+\omega_m)(\omega_q-\omega_p)}
{\sqrt{\omega_n\omega_m\omega_p\omega_q}}
\times
\big[\vec u_n^*(\vec x)\!\cdot\!\vec u_m(\vec x)\big]
\big[\vec u_p(\vec y)\!\cdot\!\vec u_q(\vec y)\big],
\label{eq:Gfinal-13}
\end{align}

\begin{align}
G_{\substack{ijkl\\nmpq}}^{(2,2)}={}&
\frac{g^2}{16\sqrt{\omega_n\omega_m\omega_p\omega_q}}
\int d^3x\,
\Big[
2\varepsilon_{\alpha ik}\varepsilon_{\alpha jl}
\big[
(\vec u_n^*\!\cdot\!\vec u_m^*)
(\vec u_p\!\cdot\!\vec u_q)
-(\vec u_n^*\!\cdot\!\vec u_q)
(\vec u_m^*\!\cdot\!\vec u_p)
\big]
\nonumber\\[-6pt]
&\hspace{3.5cm}
+\varepsilon_{\alpha ij}\varepsilon_{\alpha kl}
\big[
(\vec u_n^*\!\cdot\!\vec u_p)
(\vec u_m^*\!\cdot\!\vec u_q)
-(\vec u_n^*\!\cdot\!\vec u_q)
(\vec u_m^*\!\cdot\!\vec u_p)
\big]
\Big]
\nonumber\\[2pt]
&+\frac{g^2\varepsilon_{\alpha ij}\varepsilon_{\alpha kl}}{64\pi}
\int\frac{d^3x\,d^3y}{|\vec x-\vec y|}
\frac{(\omega_m-\omega_n)(\omega_q-\omega_p)}
{\sqrt{\omega_n\omega_m\omega_p\omega_q}}
\times
\big[\vec u_n^*(\vec x)\!\cdot\!\vec u_m^*(\vec x)\big]
\big[\vec u_p(\vec y)\!\cdot\!\vec u_q(\vec y)\big]
\nonumber\\[2pt]
&-\frac{g^2\varepsilon_{\alpha ik}\varepsilon_{\alpha jl}}{32\pi}
\int\frac{d^3x\,d^3y}{|\vec x-\vec y|}
\frac{(\omega_n+\omega_p)(\omega_m+\omega_q)}
{\sqrt{\omega_n\omega_m\omega_p\omega_q}}
\times
\big[\vec u_n^*(\vec x)\!\cdot\!\vec u_p(\vec x)\big]
\big[\vec u_m^*(\vec y)\!\cdot\!\vec u_q(\vec y)\big].
\label{eq:Gfinal-22}
\end{align}

\end{widetext}

\section{Algebraic structure and eigenstates} \label{sec:supp:singlemode_algebraic}
In this appendix, we summarize the essential commutation relations and other algebraic relations, which we use in Sec.~\ref{sec:Spectrum_of_SU2} to discuss the eigenstates of the cavity SU(2) Hamiltonian and to derive relevant matrix elements. We also refer to Refs.~\cite{Law1998,Ho1998,Koashi2000,KawaguchiUeda2012}  for similar derivations. 

\subsection{Angular momentum operators}
In the main text, we have introduced the angular momentum operators $L_i= -i \varepsilon_{ijk} a_j^\dag a_k$. We also define the bosonic modes 
\begin{align}
    a_\pm  = \frac{a_1 \mp ia_2}{\sqrt{2}},\qquad a^\dag_\pm  = \frac{a_1^\dag \pm  ia^\dag_2}{\sqrt{2}}.
\end{align}
 In terms of modes, we find that  $L_3 = a_+^\dagger a_+ - a_-^\dagger a_-$ and the raising and lowering operators, $L_\pm = L_1 \pm i L_2$, are given by
\begin{align}
L_+&= -\sqrt{2}a_+^\dagger a_3 + \sqrt{2}a_3^\dagger a_-, \\
L_- &= \sqrt{2}a_-^\dagger a_3 - \sqrt{2}a_3^\dagger a_+.
\end{align}
The operators $L_\pm$ change the angular momentum projection~$m$ and obey the usual relations 
\begin{align}
    L_\pm\ket{N,l,m}&= \sqrt{l(l+1)-m(m\pm 1)}\ket{N,l,m\pm 1}
\end{align}
and
\begin{align}\label{eq:app_general_state_Lm}
    \ket{N,l,\pm m} &= \sqrt{\frac{(l+m)!}{(2l)!(l-m)!}}L_{\mp}^{l-m}\ket{N,l,\pm l}.
\end{align}
As explained in the main text, this implies a simple expression of the $N=l=\pm m$ eigenstate,
\begin{align}\label{eq:app_max_l_state}
    \ket{l,l,\pm l}&=\frac{(\pm 1)^l}{\sqrt{l!}}a_\pm ^{\dagger l}\ket{\vac},
\end{align}
where the global sign has been chosen to be consistent with the convention for the general eigenstates discussed below. 

\subsection{Glueball operators}
In the main text, we have introduced the glueball raising and lowering operators, 
\begin{align}
    B^\dagger &= \frac{a_1^{\dagger2} + a_2^{\dagger2} + a_3^{\dagger2}}{\sqrt{6}}, \qquad B =  \frac{a_1^2 + a_2^2 + a_3^2}{\sqrt{6}}.
\end{align}
From explicit calculations, we obtain the commutators $[B, B^\dagger]= 1 + 2N/3$ and $[B, L_i] = [B, L^2] = 0$, together with 
\begin{align}
    B^\dagger B &= \frac{1}{6}\left[N(N+1) - L^2\right], \\
    B B^\dagger &= \frac{1}{6}\left[(N+2)(N+3) - L^2\right].
\end{align}
From the properties above, we find that the glueball ladder operators change $N$ by two without changing $l$ and $m$,
\begin{align}
    B^\dagger\ket{N,l,m} &= \sqrt{\frac{(N+2)(N+3)-l(l+1)}{6}}\ket{N+2,l,m}, \label{eq:app_Bdag_general}\\
    B\ket{N,l,m} &= \sqrt{\frac{N(N+1)-l(l+1)}{6}}\ket{N-2,l,m}. \label{eq:app_B_general}
\end{align}
Specifically, when we apply these relations to glueball states $\ket{\mathcal{N}}_g\equiv\ket{2\mathcal{N},0,0}$, we obtain the ladder relations
\begin{align}
    B^\dagger \ket{\mathcal{N}}_g &= \sqrt{\frac{(2\mathcal{N}+2)(2\mathcal{N}+3)}{6}}\ket{\mathcal{N}+1}_g, \\
    B \ket{\mathcal{N}}_g &= \sqrt{\frac{2\mathcal{N}(2\mathcal{N}+1)}{6}}\ket{\mathcal{N}-1}_g,
\end{align}
together with the following normalized expressions for the glueball states,
\begin{align}
\ket{\mathcal{N}}_g &= \frac{B^{\dagger \mathcal{N}}}{\sqrt{C_\mathcal{N}}}\ket{\vac},\qquad
    C_\mathcal{N}= \frac{(2\mathcal{N}+1)!}{6^\mathcal{N}}.
\end{align}

\subsection{Eigenstates and normalization factors}

We can now combine the results from the previous subsection to derive the full expression for the general eigenstates in Eq.~\eqref{eq:general_eigenstate_exp}. We first combine Eq.~\eqref{eq:app_general_state_Lm} and Eq.~\eqref{eq:app_max_l_state} to obtain
\begin{equation}
|l,l,m\rangle = \sqrt{\frac{(l+m)!}{l!(2l)!(l-m)!}}L_{-}^{l-m} (a_+^\dag)^l\ket{{\rm vac}}.
\end{equation}
We then use Eq.~\eqref{eq:app_Bdag_general} and Eq.~\eqref{eq:app_B_general}  together with  the factorization $x(x+1)-y(y+1)=(x-y)(x+y+1)$, to obtain a relation between a maximal angular momentum eigenstate~$\ket{l,l,m}$ and its corresponding eigenstate~$\ket{l+2\mathcal{N},l,m}$ containing $\mathcal{N}$ extra glueballs. These relations read 
\begin{align}
    \ket{l+2\mathcal{N},l,m} &= d_{\mathcal{N},l}B^{\dagger \mathcal{N}}\ket{l,l,m}, \label{eq:app:Bdag_action}\\
    \ket{l,l,m} &= d_{\mathcal{N},l}B^{ \mathcal{N}}\ket{l+2\mathcal{N},l,m},
\end{align}
where the prefactor is given by
\begin{align}
    d_{\mathcal{N},l} &= \sqrt{\frac{(2l+1)!!6^\mathcal{N}}{ (2\mathcal{N})!!(2l+2\mathcal{N}+1)!!}} \\
    &=\sqrt{\frac{(2l+2)!(l+\mathcal{N}+1)!6^\mathcal{N}}{ (2l+2\mathcal{N}+2)!(l+1)!\mathcal{N}!}}.
\end{align}
Combining the above results, we obtain the general expression~\eqref{eq:general_eigenstate_exp} for an arbitrary eigenstate
\begin{align}\label{eq:app:eigenstate_from+}
    \ket{l+2\mathcal{N},l,  m} &= c_{\mathcal{N},l,m} B^{\dagger \mathcal{N}}L_{-}^{l-m}a_{+}^{\dagger l}\ket{\vac},
\end{align}
with the prefactor
\begin{align}
    c_{\mathcal{N},l,m} &= \sqrt{\frac{(l+m)!(2l+2)!(l+\mathcal{N}+1)!6^\mathcal{N}}{l!(2l)!(l-m)!(2l+2\mathcal{N}+2)!(l+1)!\mathcal{N}!}}.
\end{align}
This expression is particularly convenient for evaluating states with $0 \leq m\lesssim l$. In the opposite range, for states with $0>m\gtrsim -l$, it is convenient to use the similar expression obtained from the mode $a_-^\dagger$,
\begin{align}\label{eq:app:eigenstate_from-}
    \ket{l+2\mathcal{N},l,-m} &= (-1)^l c_{\mathcal{N},l,m} B^{\dagger \mathcal{N}}L_{+}^{l-m}a_{-}^{\dagger l}\ket{\vac},
\end{align}
where the global sign has been fixed such that the two expressions~\eqref{eq:app:eigenstate_from+} and~\eqref{eq:app:eigenstate_from-} lead to the same phase convention to the eigenstate~$\ket{N,l,m}$.

\subsection{Matrix elements}
As explained in the main text, the bosonic operator $a_3^\dagger$ increases $N$ by one, changes $l$ by one, but does not change $m$. Here we evaluate the corresponding matrix elements
\begin{align}
    \beta_{N,l,m}^{(\pm)} = \bra{N+1,l\pm1,m}a_3^\dagger \ket{N,l,m}.
\end{align}
To do so, we first obtain a relation between the matrix elements for $N=l$ and $N=l+2\mathcal{N}$.  Using Eq.~\eqref{eq:app:Bdag_action}, we obtain
\begin{align}
    \beta_{l+2\mathcal{N},l,m}^{(+)} &= \frac{d_{\mathcal{N},l}}{d_{\mathcal{N},l+1}}\beta_{l,l,m}^{(+)} = \sqrt{\frac{2l+2\mathcal{N}+3}{2l+3}}\beta_{l,l,m}^{(+)},\\
    \beta_{l+2\mathcal{N},l,m}^{(-)} &= \frac{d_{\mathcal{N},l}d_{1,l-1}}{d_{\mathcal{N}+1,l-1}}\beta_{l,l,m}^{(-)}=\sqrt{\mathcal{N}+1}\beta_{l,l,m}^{(-)}.
\end{align}
To obtain the values for $\beta_{l,l,m}^{(\pm)}$, we explicitly expand the states using Eq.~\eqref{eq:app:eigenstate_from+},
\begin{align}
    \beta_{l,l,m}^{(+)}&=c_{0,l+1,m}c_{0,l,m}\bra{\vac}a_+^{l+1}L_+^{l+1-m}a_3^\dagger L_-^{l-m}a_+^{\dagger l}\ket{\vac},\\
    \beta_{l,l,m}^{(-)}&=c_{1,l-1,m}c_{0,l,m}\bra{\vac}a_+^{l-1}L_+^{l-1-m}Ba_3^\dagger L_-^{l-m}a_+^{\dagger l}\ket{\vac}.
\end{align}
For $\beta_{l,l,m}^{(+)}$, we use the commutation relations
\begin{align}
    [L_+,a_3^\dagger]=-\sqrt{2}a_+^\dagger \quad;\quad [L_+,a_+^\dagger]=0,
\end{align}
such that
\begin{align}
    [L_+^q,a_3^\dagger]=-q\sqrt{2}a_+^\dagger L_+^{q-1}.
\end{align}
Combined with $[a_+,a_3^\dagger]=0$ and $[a_+^{l+1},a_+^\dagger]=(l+1)a_+^l$, as well as Eq.~\eqref{eq:app:eigenstate_from+}, we obtain
\begin{align}
    \beta_{l,l,m}^{(+)}&=-\sqrt{2}(l+1)(l-m+1)\frac{c_{0,l+1,m}}{c_{0,l,m}},\\
    &=-\sqrt{\frac{(l+m+1)(l-m+1)}{2l+1}}.
\end{align}
For $\beta_{l,l,m}^{(-)}$, we use $[B,a_3^\dagger]=\frac{2}{\sqrt{6}}a_3$ and $B L_-^{l-m}a_+^{\dagger l}\ket{\vac}=0$, since $L_-^{l-m}a_+^{\dagger l}\ket{\vac}$ is a maximal-$l$ state. We obtain
\begin{align}
    \beta_{l,l,m}^{(-)}&= \frac{2c_{1,l-1,m}}{\sqrt{6}c_{0,l-1,m}}\left(\beta_{l-1,l-1,m}^{(+)}\right)^*,\\
    &= -\sqrt{\frac{2(l+m)(l-m)}{(2l+1)(2l-1)}}.
\end{align}

Combining these results, we obtain the general expression of the non-zero matrix elements of $a_3^\dagger$,
\begin{align}
    \beta_{N,l,m}^{(+)}&=-\sqrt{\frac{(N+l+3)(l+m+1)(l-m+1)}{(2l+3)(2l+1)}},\\
    \beta_{N,l,m}^{(-)}&=-\sqrt{\frac{(N-l+2)(l+m)(l-m)}{(2l+1)(2l-1)}}.
\end{align}
Note that these expressions are real, such that the matrix elements of $a_3$ are simply given by 
\begin{align}
    \bra{N-1,l\pm1,m}a_3 \ket{N,l,m} = \beta_{N-1,l\pm 1,m}^{(\mp)}.
\end{align}

\section{Perturbation theory for a driven cavity}\label{app_PerturbationTheory}
In Sec.~\ref{sec:DrivenCavity}, we considered the case of a weakly-driven cavity with $\varepsilon,|\Delta| \ll U$. In this limit, the sector of glueball states $|\mathcal{N}\rangle_g=|2\mathcal{N},0,0\rangle$ with energies $-2\Delta \mathcal{N}$ are energetically separated from the $l=1$ states, with an energy $\sim 2U$. 
Therefore, we can treat $H_\varepsilon=\varepsilon(a_3+a_3^\dag)$ as a small perturbation and derive an effective Hamiltonian $H_{\rm eff}= -\Delta N +H^{(2)}$, up to second order in $\varepsilon$. Here,  
\begin{equation}
H^{(2)}= -\sum_{\mathcal{N},\mathcal{N}^\prime} |\mathcal{N}^\prime\rangle_g\frac{\gluebra{\mathcal{N}^\prime} H_\varepsilon \mathbbm{P}_{1}H_\varepsilon|\mathcal{N}\rangle_g}{2U}\gluebra{\mathcal{N}}
\end{equation}
and 
\begin{equation}
\mathbbm{P}_{1}= \sum_{\mathcal{N}} |2\mathcal{N}+1,1,0\rangle \langle 2\mathcal{N}+1,1,0|
\end{equation}
is the projector on the excited subspace. Since $\mathbbm{P}_{1}a_3^\dag \ket{\mathcal{N}}_g= a_3^\dag \ket{\mathcal{N}}_g$, etc., the projector can be omitted and we are left with two nontrivial matrix elements. The diagonal element, 
\begin{align}
\gluebra{\mathcal{N}^\prime} (2a_3^\dag a_3+1) |\mathcal{N}\rangle_g =\left( \frac{4\mathcal{N}}{3}+1\right)\delta_{\mathcal{N},\mathcal{N}^\prime},
\end{align}
can be obtained directly using the relation $a_3^\dag a_3=N/3$ within the glueball subspace. It translates into an energy shift $-\varepsilon^2 2\mathcal{N}/(3U)$ for the glueball states.

The second relevant matrix element is 
\begin{align}
\gluebra{\mathcal{N}^\prime}  a_3^{2\dag} |\mathcal{N}\rangle_g&= \beta_{2\mathcal{N},0,0}^{+}\beta_{2\mathcal{N}+2,1,0}^{-}\delta_{\mathcal{N}+1,\mathcal{N}^\prime}\\
&=\sqrt{ \frac{(2\mathcal{N}+3)(2\mathcal{N}+2)}{9}}\delta_{\mathcal{N}+1,\mathcal{N}^\prime}.
\end{align}
It multiplies terms $\sim \ket{\mathcal{N}+1}_g\bra{\mathcal{N}}$  in $H^{(2)}$. In view of the relation
\begin{equation}
\ket{\mathcal{N}+1}_g\bra{\mathcal{N}} =\sqrt{ \frac{6}{(2\mathcal{N}+3)(2\mathcal{N}+2)}} B^\dag \ket{\mathcal{N}}_g\bra{\mathcal{N}},
\end{equation}
these contributions can be expressed in term of the glueball raising operator only. Overall we obtain the effective Hamiltonian in Eq.~\eqref{eq:Heff_drive}.

\section{Effective Hamiltonian for a cavity array}\label{sec:array_Heff}
In this appendix, we outline the derivation of the effective low-energy Hamiltonian $H_{\rm eff}$ in Eq.~\eqref{eq:Heff}, starting from the Hamiltonian $H_{\rm array}$ in Eq.~\eqref{eq:Harray} and assuming $J\ll U$. We change into an interaction picture with respect to $H_0=\sum_X \omega_0 N_X$ and write the full Hamiltonian for the cavity array as 
\begin{align}
    H_{\rm array} = H_{\rm cav} + V_J + V_C, 
\end{align}
where $H_{\rm cav} = \cavcoupling\sum_X L_X^2$ and 
\begin{align}
    V_J &= - J \sum_{X} \sum_{i=1}^3 \left( a_{i,X}^\dagger a_{i,X+1}+ {\rm H.c.}\right)
\end{align}
is the tunneling Hamiltonian, which we treat as a perturbation. The last term, $V_C=(1/2)\sum_{X\neq Y} \tilde U_C(X,Y)L_{X,i}L_{Y,i}$, accounts for the long-range part of the Coulomb interaction. This contribution vanishes within the glueball sector and for $\tilde U_C\lesssim  2U$ it only modifies the energy denominator in the final result. Therefore, we neglect this term in the following derivation.    

In the following we denote by $\mathbbm{P}$ the projector on the glueball states with quantum numbers $l_X=0$, and $\mathbbm{Q}=\mathbbm{1}-\mathbbm{P}$. In the limit $J\rightarrow0$, we find that $\mathbbm{P}H_{\rm cav}\mathbbm{P}=0$ and up to second order in $J/U$, the effective Hamiltonian is given by 
\begin{align}
    H_{\rm eff} = - \mathbbm{P}V_J \mathbbm{Q} \frac{1}{H_{\rm cav}}\mathbbm{Q}V_J\mathbbm{P},
\end{align}
To evaluate this term, we note that 
\begin{equation}
\mathbbm{Q} \frac{1}{H_{\rm cav}}\mathbbm{Q}(a^\dag_{i,X} a_{i,X+1}) \mathbbm{P}= -\frac{1}{4U}(a^\dag_{i,X} a_{i,X+1}) \mathbbm{P},
\end{equation}
because $a^\dag_{i,X}$ and $a_{i,X+1}$ create excitations with $l_{X}=1$ and $l_{X+1}=1$ with energy $2U$ in neighboring cavities. Therefore,
\begin{align}\label{eq:Heff_step1}
\begin{split}
    H_{\rm eff} = -\frac{J^2}{4U}  \sum_{X} \sum_{i,j}& \mathbbm{P}  
\left( a_{i,X}^\dagger a_{i,X+1}+ {\rm H.c.}\right)\\
\times& \left( a_{j,X}^\dagger a_{j,X+1}+ {\rm H.c.}\right) \mathbbm{P},
\end{split}
\end{align}
where we have used that $\mathbbm{P}a_{i,X}^\dagger a_{i,X+1}a_{i,Y}^\dagger a_{i,Y+1}\mathbbm{P}=0$, etc. when $X\neq Y$.

To evaluate this expression, we consider a single pair of sites, for which the product of the two brackets in Eq.~\eqref{eq:Heff_step1} reads 
\begin{align}
\begin{split}
(...)(...) =\,&  2 a_{i,X}^\dag a_{j,X}  a^\dag_{i,X+1} a_{j,X+1}\\ 
&+\left(a_{i,X}^\dag a_{j,X}^\dag  a_{i,X+1} a_{j,X+1} + {\rm H.c.}\right)\\
&+ \left( a_{i,X}^\dag a_{i,X} + a^\dag_{i,X+1} a_{i,X+1}\right) \delta_{ij}.
\end{split}
\end{align}
To proceed we make use of the relations
\begin{align}
\mathbbm{P} a_i^\dag a_j^\dag \mathbbm{P} = \delta_{ij}\sqrt{\frac{2}{3}} B^\dag, \qquad  \mathbbm{P} a_i^\dag a_j \mathbbm{P} =  \delta_{ij}\frac{2}{3} \mathcal{N}.
\end{align}
They can be derived by keeping in mind that any product $a^\dag_i a_j$ or $a^\dag_i a^\dag_j$ with $i\neq j$ creates unpaired bosons that are outside the glueball subspace. The diagonal terms are symmetric in the flavor index and can thus be evaluated for $a_3^{2\dag}$ and $a_3^\dag a_3$, similar as in Sec.~\ref{app_PerturbationTheory} above. As a result, we obtain the effective Hamiltonian in Eq.~\eqref{eq:Heff}.

\section{Circuit model}\label{app:circuit}

\begin{figure}
    \centering
    \includegraphics[width=\linewidth]{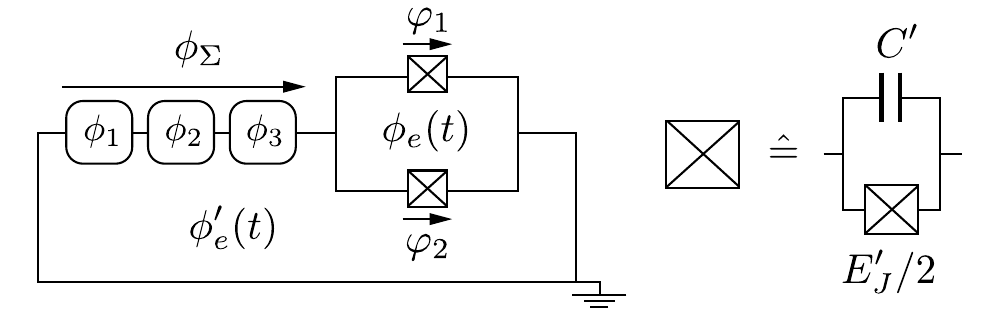}
    \caption{Sketch of the circuit layout and the relevant phase variables and external fluxes assumed in the derivation of the circuit Hamiltonian in Eq.~\eqref{eq:Hcircuit1}. }
    \label{fig:circuit_quantization}
\end{figure}

In this appendix, we present additional details about the derivation of the circuit Hamiltonian in Eq.~\eqref{eq:Hcircuit1}. The corresponding circuit layout is sketched in Fig.~\ref{fig:circuit_quantization} and is described by the Lagrangian
\begin{equation}
\mathcal{L}= \sum_i \mathcal{L}_i + \mathcal{L}_a.
\end{equation}
Here, the Lagrangians $\mathcal{L}_i\equiv \mathcal{L}_i(\phi_i,\dot\phi_i)$ model the three subcircuits for the gauge-boson modes. Each of them represents a fluxonium-type circuit~\cite{Manucharyan2009} and can be quantized in the usual way~\cite{VoolDevoret2017,Krantz2019}. The additional Lagrangian,
\begin{equation}
\mathcal{L}_a= \mathcal{T}_a-\mathcal{V}_a = \sum_{\eta=1,2}\frac{C^\prime_\eta}{2} \dot \varphi_\eta^2 + \frac{E_J^\prime}{2} \cos(\varphi_\eta),
\end{equation}
accounts for the kinetic and potential energy of the auxiliary coupling circuits, with $\varphi_\eta$ denoting the phases across the upper and lower junctions. These phase variables are not independent degrees of freedom. They are fixed by the flux quantization conditions in the two closed loops, 
\begin{align}
\varphi_1-\varphi_2 = \phi_{e}(t),\qquad {\rm and}\qquad \phi_\Sigma+\varphi_2&= \phi^\prime_{e}(t),
\end{align}
where $\phi_\Sigma=\sum_i \phi_i$. The external fluxes are applied at a fixed ratio 
\begin{align}
\phi^\prime_e(t)= -\phi_e(t)/2,
\end{align}
such that 
\begin{align}
\varphi_{1,2} = - \phi_\Sigma \pm \phi_e(t)/2.
\end{align}
This choice ensures that in the resulting kinetic term,
\begin{equation}
\mathcal{T}_a= \frac{(C_1^\prime+C_2^\prime)}{2} \dot{\phi}^2_{\Sigma}- \left(\frac{C_1^\prime}
{2}-\frac{C_2^\prime}
{2}\right)\dot{\phi}_{\Sigma} \dot \phi_e, 
\end{equation}
the driving term $\sim \dot{\phi}_\Sigma$ vanishes also for an oscillating flux when the junctions are symmetric, $C_1^\prime\simeq C_2^\prime$. The static correction to the capacitive energy can be neglected when $C^\prime$ is small compared to the capacitance of the subcircuits. For the remaining potential energy of the coupler we obtain 
\begin{equation}
\begin{split}
\mathcal{V}_a&=-\frac{E^\prime_J}{2} \cos\left( \phi_\Sigma -\frac{\phi_{e}(t)}{2} \right) -\frac{E^\prime_J}{2} \cos\left( \phi_\Sigma +\frac{\phi_{e}(t)}{2} \right)\\
&= - E^\prime_J \cos\left(\phi_\Sigma\right) \cos\left(\frac{\phi_{e}(t)}{2}\right).
\end{split}
\end{equation}

\bibliography{refs}

\end{document}